\documentclass[aip,reprint,pop]{revtex4-1}
\usepackage{amsmath,amssymb,graphicx}
\draft 

\begin{document}
\preprint{LA-UR-19-28713}

\title{Experimental characterization of a section of a spherically imploding plasma liner formed by merging
hypersonic plasma jets} 



\author{K. C. Yates}
\email[]{kyates@lanl.gov}
\affiliation{Physics Division, Los Alamos National Laboratory, Los Alamos, NM  87545, USA}
\affiliation{Electrical and Computer Engineering Department,
University of New Mexico, Albuquerque, NM 87131, USA }
\author{S. J. Langendorf}
\email[]{samuel.langendorf@lanl.gov}
\author{S. C. Hsu}
\email[]{scotthsu@lanl.gov}
\author{J. P. Dunn}
\affiliation{Physics Division, Los Alamos National Laboratory, Los Alamos, NM  87545, USA}
\author{S. Brockington}
\author{A. Case}
\author{E. Cruz}
\author{F. D. Witherspoon}
\affiliation{HyperV Technologies Corp., Chantilly, VA 20151, USA}
\affiliation{HyperJet Fusion Corporation, Chantilly, VA 20151, USA}
\author{Y. C. F. Thio}
\affiliation{HyperJet Fusion Corporation, Chantilly, VA 20151, USA}
\author{J. T. Cassibry}
\affiliation{Propulsion Research Center, University of Alabama in Huntsville, Huntsville, AL 35899, USA}
\author{K. Schillo}
\affiliation{Propulsion Research Center, University of Alabama in Huntsville, Huntsville, AL 35899, USA}
\author{M. Gilmore}
\affiliation{Electrical and Computer Engineering Department,
University of New Mexico, Albuquerque, NM 87131, USA }

\date{\today}

\begin{abstract}

We report experimental results on merging of hypersonic plasma jets, which
is the fundamental building block for forming spherically imploding plasma liners as a potential standoff
compression driver for magneto-inertial fusion. Jets are formed and launched by contoured-gap coaxial
plasma guns mounted at the six vertices and the center of a hexagon covering approximately one-tenth of
the surface area of a 9-ft.-diameter spherical chamber. First, from experiments with two and three
merging jets of four different species (N, Ar, Kr, Xe), we show that (1) density spatial non-uniformities can be
large (with electron-density jumps ranging from 2.9 for N to 6.6 for Xe) when shocks form upon jet merging,
but smaller (density jumps $<$2) when shocks do not form; (2) jet impurities (20\% Ti in these experiments)
can increase the level of density spatial non-uniformity by increasing the collisionality of jet merging, leading
to shock formation rather than potentially more desirable shockless jet merging; and (3) the liner Mach
number can remain high ($\gtrsim 10$), as required for plasma liners to be an effective compression driver. Second,
from experiments with six and seven merging jets using Ar, we present results with improved jet-to-jet
mass balance of $<$2\% across jets, including (1) evidence of substantially increased balance in
the jet merging and symmetry of the liner structure, and (2) potentially favorable changes in the jet-merging
morphology with the addition of the seventh jet. For both experiments, we
present comparisons between experimental and synthetic data from three-dimensional hydrodynamic codes.
\end{abstract}

\pacs{}

\maketitle 

\section{Introduction}
\label{sec:intro}

Magneto-inertial fusion (MIF), aka magnetized target fusion (MTF),
is a class of pulsed fusion approaches in which an imploding liner compresses
a magnetized target plasma to fusion conditions, \cite{lindemuth83, kirkpatrick95, lindemuth15, wurden16}
at ion densities intermediate between those of magnetic and inertial fusion.
Many MIF embodiments have been pursued over a period spanning more than forty years, e.g.,
the development of rotating cylindrical liquid liners\cite{turchi80,turchi17} intended to compress 
a field-reversed configuration, MAGnitnoye Obzhatiye (MAGO) or 
magnetic compression,\cite{garanin1998mago} cylindrical solid-liner compression of an FRC,\cite{intrator04,degnan2013recent}
acoustically driven liquid-liner compression of a spherical tokamak, \cite{laberge2019magnetized} and 
magnetized liner inertial fusion\cite{slutz2010pulsed,gomez2014experimental} (MagLIF, which is a cylindrical
solid-liner compression of a laser-preheated magnetized plasma).   MagLIF provided 
a definitive demonstration of proof-of-concept for MIF by achieving multi-keV temperatures\cite{gomez2014experimental}
and $BR$ (product of magnetic field times fuel radius) values\cite{schmit14} approaching those needed
for fuel self-heating from energy deposition by fusion-produced $\alpha$ particles.\cite{basko2000ignition}
To meet the economic requirements of a power plant, it may be necessary for an MIF
embodiment to have high repetition rate (e.g., $\sim 1$~Hz) and low cost per shot
(e.g., few cents/shot amortized over the life of the power plant).  This tends to favor liquid and plasma
liners, which avoid the repetitive mass destruction associated with solid liners that lead
to lower repetition rate and higher cost per shot.

Since 2009, a multi-institutional collaboration led by Los Alamos National Laboratory (LANL) has been exploring the 
development of a high-shot-rate, low cost-per-shot compression driver for MIF based on the concept of plasma-jet-driven MIF, or
PJMIF,\cite{thio99,thio01,parks08pop,santarius2012compression,hsu2012spherically,knapp2014possible,langendorf2017semi,thio2019plasma}
in which a spherically imploding plasma liner is formed via merging hypersonic plasma jets.  The
chief advantages of PJMIF are (1)~high implosion speeds ($>50$~km/s) to overcome
the rate of energy loss in the magnetized plasma target and (2)~several-meter
standoff of the plasma-formation hardware (plasma guns)
to allow for reasonably long periods between maintenance or replacement in a power plant.  The key
disadvantages of PJMIF are the early stage of development and challenges in forming both the spherically imploding plasma 
liner\cite{hsu2012spherically,hsu2018physics} and a compatible magnetized plasma target.\cite{hsu2019magnetized}
Furthermore, merging plasma jets will seed non-uniformities in the
liner that could lead to intolerable levels of Rayleigh-Taylor instability and liner/fuel mix when the
heavier liner decelerates against the lighter target plasma.  The amount of
liner non-uniformity resulting from merging plasma jets and the amount that can be tolerated for PJMIF to 
remain viable are both open research questions.  Plasma liners may also have limited hydrodynamic efficiency
in transferring energy to the target due in part to radial expansion of the liner during implosion.\cite{langendorf2017semi}

A body of published literature lays out the theoretical issues and numerical scaling-studies
of plasma-liner formation and implosion via merging hypersonic plasma jets.\cite{thio99,cassibry2009estimates,samulyak2010spherically,awe2011one,hsu2012spherically,santarius2012compression,cassibry2012tendency,davis2012one,cassibry2013ideal,kim2013structure,knapp2014possible,hsu2018physics,langendorf2017semi}
Previous research on the LANL Plasma Liner Experiment (PLX),\cite{hsu2012spherically,hsu2015laboratory} using parallel-plate
mini-railguns developed by HyperV Technologies
Corp.,\cite{witherspoon2011development,brockington2012hyperv,case2013merging,messer2013nonlinear}
studied single plasma-jet propagation,\cite{hsu2012experimental} two-jet oblique
merging,\cite{merritt2013experimental,merritt2014experimental} and two-jet head-on 
merging.\cite{moser2015experimental}  These earlier experiments led to the identification and characterization of plasma shock 
formation between merging jets
that was shown to be consistent with hydrodynamic oblique shock theory.\cite{merritt2013experimental,merritt2014experimental}   
The merging of head-on plasma jets was shown to lead to slightly higher electron temperature $T_e$ and
mean-charge-state $\bar{Z}$, tending to
reduce the counter-streaming ion--ion mean free path and increasing the collisionality of the jet
merging.\cite{moser2015experimental} 

The aim of this work is to characterize in detail the merging of up to seven hypersonic plasma jets
launched by new contoured-gap coaxial guns\cite{hsu2018experiment,thio2019plasma} developed under
the ARPA-E ALPHA program.\cite{nehl19jfe} 
First, we report results from two- and three-jet merging experiments repeated for different
gas species (N, Ar, Kr, Xe), augmenting recent shock-ion-heating 
studies,\cite{langendorf2018experimental,langendorf2019experimental} which are based on some of the same experimental
campaigns presented here.  As mentioned earlier, plasma-jet merging generates non-uniformities upon formation of the 
imploding plasma liner, leading to strong, localized ion
heating\cite{langendorf2018experimental} and an impulsive increase in the ion sound speed $C_s$ and 
decrease in Mach number $M$.  This degrades the liner's ability to effectively compress a magnetized fusion target.  This
work provides experimental data on the plasma parameters and new understanding of shock-formation dynamics and
evolution of $M$ during two- and three-jet merging.
Second, we evaluate the symmetry in a section of the spherically imploding plasma liner 
that is formed by merging six and seven hypersonic plasma jets with reduced ($<2$\%) jet-to-jet mass variation 
across all the jets, as compared to earlier results with $>20$\% mass variation across six jets.\cite{hsu2018experiment}  
While merging two and three
plasma jets is the most fundamental building block for
forming spherically imploding plasma liners via merging plasma jets, the merging of six and seven
jets covering approximately one-tenth of the surface area of a sphere is the next step toward evaluating the macro-structure 
and morphology of a section of the plasma liner. The data presented include time- and space-resolved
electron density $n_e$ and $T_e$ measurements from interferometry and survey spectroscopy, respectively,
and end-on images of six- and seven-jet merging.  All of the data presented in this paper
are being used
to benchmark multi-physics models and codes\cite{schillo2019suite,shih2019simulation} to further evaluate 
plasma-jet merging, plasma-liner formation, and the PJMIF concept.\cite{thio99,hsu2012spherically,thio2019plasma}

The paper is organized as follows.  Section~\ref{sec:setup} describes
the experimental and diagnostic setups for both the two/three and six/seven jet-merging experiments.
Sections~\ref{sec:results1} and \ref{sec:results2} present results and discussion of the two/three and six/seven
jet-merging experiments, respectively.  Section~\ref{sec:summary} provides a summary and conclusions.
Appendix~\ref{sec:more_data} includes the N, Kr, and Xe data associated with Sec.~\ref{sec:results1}, and
Appendix~\ref{sec:jets} presents details on the improved jet-to-jet balance associated with Sec.~\ref{sec:results2}.

\section{Experimental setup}
\label{sec:setup}

\subsection{PLX facility and diagnostics}

Many details about PLX, on which the present experiments were performed, have been presented
elsewhere.\cite{hsu2012experimental,hsu2015laboratory,hsu2018experiment,langendorf2019experimental,thio2019plasma}  
Here, we summarize the particular 
details pertinent to this paper.
PLX has a 9-ft.-diameter, stainless-steel spherical vacuum chamber.  For this work, 
seven coaxial plasma guns are
mounted on the chamber in a hexagonal pattern (with six guns on the vertices and one in the middle), 
covering roughly one-tenth of the surface area of the spherical vacuum chamber.  

\begin{figure}[!tb]
\includegraphics[width=3.2truein]{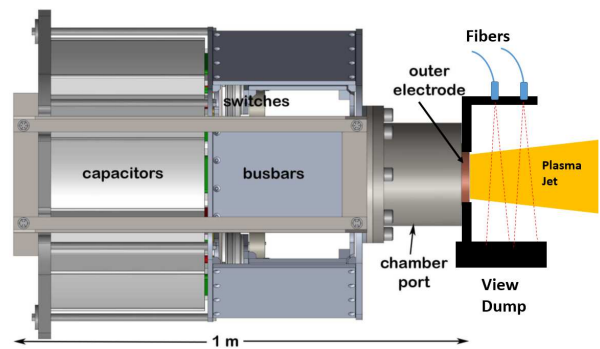}
\caption{\label{fig:gun} The coaxial plasma gun used in these experiments, with an integrated -5-kV (negative
polarity on the inner electrode), 575-$\mu$F capacitor bank driving the
gun electrodes.  The timing and speed 
of each plasma jet are measured by fiber-coupled photodiodes with viewing sightlines separated by 2~cm as shown.}
\end{figure}

The coaxial plasma guns (Fig.~\ref{fig:gun}) are designed and built
by HyperV Technologies Corp.\ and described in detail elsewhere.\cite{hsu2018experiment}  
In these experiments, initial plasma-jet densities are $n_i \sim 10^{16}$~cm$^{-3}$, $T_e \approx 1.5$~eV, 
$\bar{Z}\approx 1$--2, speed $v_{\rm jet} \approx 50$~km/s, and
$M=v_{\rm jet}/C_s \gtrsim 10$.  
The banks driving each set of gun electrodes and all the gas valves (GV), pre-ionization (PI), and
electrode-bank master-trigger (MT) systems used peak
voltages of -4.5, 8.5, 24, and -28~kV, respectively, with capacitances of 575, 96, 6, and 6~$\mu$F, respectively (note
that there is a separate electrode bank for each gun, but all seven guns share the GV, PI, and MT banks). 
The GV capacitor
bank is triggered at $t= -600$ to -300~$\mu$s, PI bank triggered at $t=-20$~$\mu$s, and gun-electrode banks
at $t=0$. 

The diagnostic suite includes a twelve-chord interferometer, visible survey spectrometer, high-resolution Doppler
spectrometer, photodiode array at each gun nozzle to measure jet speed, and a single-frame camera with an
intensified charge-coupled-device (iCCD) detector  The interferometer uses a 320-mW,
651-nm solid-state laser in a heterodyne configuration, 
which is an upgrade from the previous eight-chord system.\cite{merritt2012multi1,merritt2012multi2}  Each
probe beam is approximately 0.3~cm in diameter, which sets the spatial resolution transverse to each chord.  Visible survey
spectroscopy is fielded with 0.160-nm/pixel resolution at 510~nm and coupled to a PI-MAX2 intensified charged-couple-device
(iCCD) camera, 
with a typical exposure of 1--2~$\mu$s and a field-of-view of 1--2~cm in diameter.  High-resolution spectroscopy is fielded using 
a 4-m McPherson 2062DP, with 2400-mm\textsuperscript{-1} grating and 1.52~pm/pixel at 480.6~nm.  The 4-m spectrometer is 
coupled to a Stanford Computer Optics 4 Quik E iCCD with typical exposure time of 1~$\mu$s and a field-of-view of 1--2~cm
in diameter.  The survey and high-resolution spectrometer viewing optics (one chord each) are moved when necessary
to different positions between shots. Two light-collecting optical fibers separated by 2~cm were installed near the exit of each gun 
as shown in Fig.~\ref{fig:gun}.  Light is collected through a 1-mm, 5/16-in.-deep pinhole with SH-4001 fibers and relayed to a 
100-MHz digitizer board, which provides jet velocity via time-of-flight analysis.  Single-frame iCCD images are obtained with a 
PCO DiCam Pro camera ($1280\times1024$ pixels), with exposure time of 10~ns.  All experimental times $t$ are given relative 
to the trigger time of the gun electrodes.  Details of the data reduction and analysis methodologies
have been used extensively and reported 
elsewhere.\cite{hsu2012experimental,merritt2014experimental,moser2015experimental,langendorf2019experimental}

\subsection{Setup for merging of two and three plasma jets}

\begin{figure}[htb!]
\includegraphics[width=3.0truein]{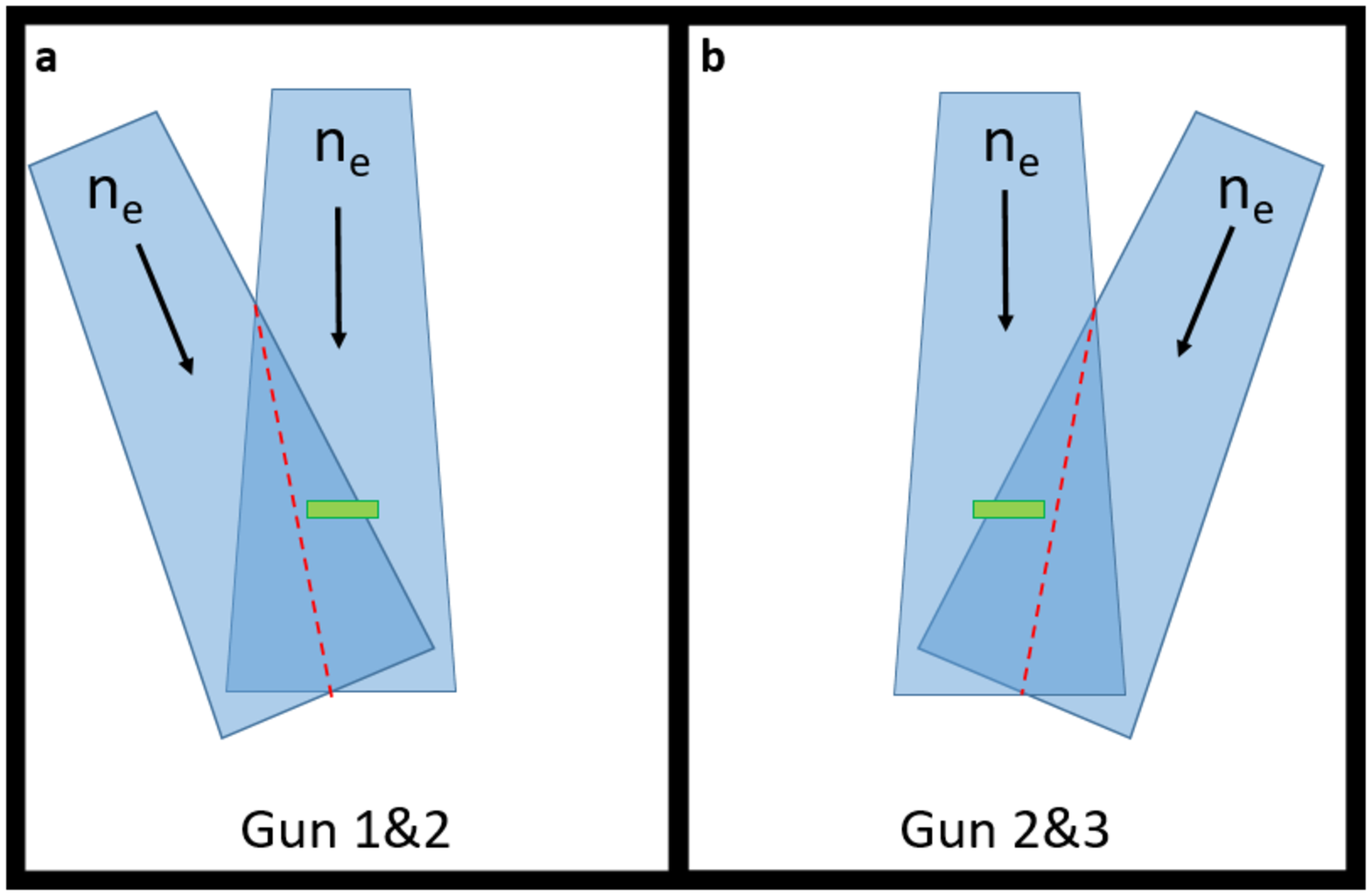}
\includegraphics[width=3.0truein]{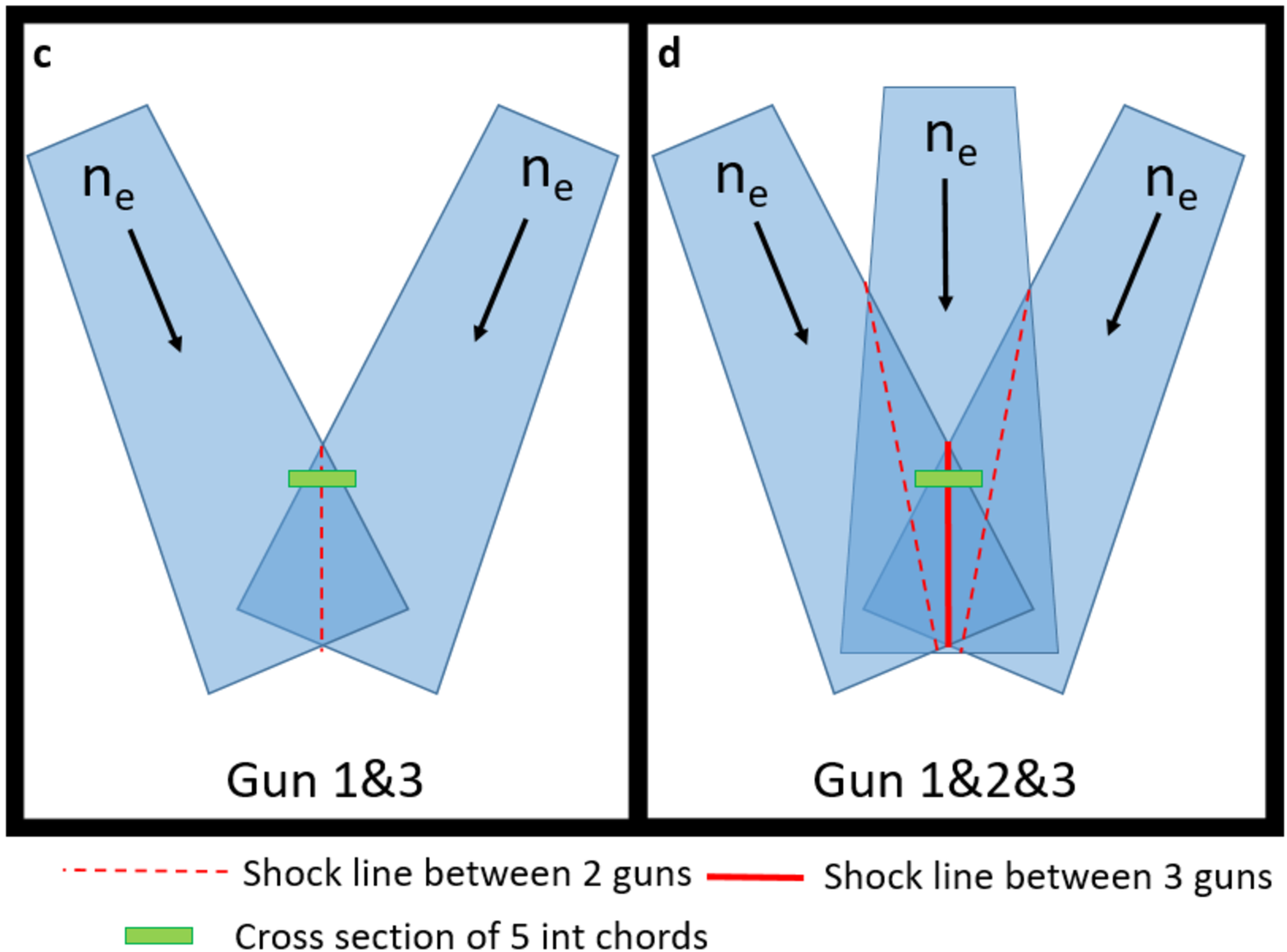}
\caption{\label{fig:cartoon} Setup for results of Sec.~\ref{sec:results1}: merging geometry for two- and three-jet merging experiments.  
The angle is 23.2$^\circ$ between
guns 1,2 and 2,3, and 41$^\circ$ between guns 1,3.  See Fig.~\ref{fig:6gundiag2} for gun positions.}
\end{figure}

\begin{figure}
\includegraphics[width=3.2truein]{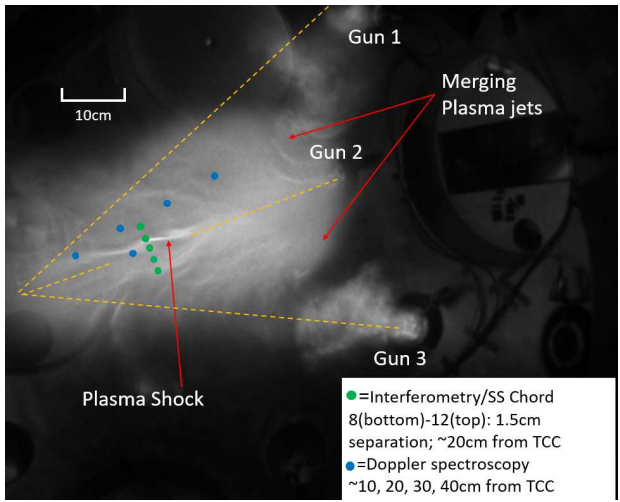}
\caption{\label{fig:diagnostics}
Setup for results of Sec.~\ref{sec:results1}:  annotated,
side-on visible image (10-ns exposure, $t=36$~$\mu$s) 
of three merging argon plasma jets.  Shown are
the locations of five interferometry and
survey-spectroscopy viewing chords (green dots)
and five high-resolution Doppler spectroscopy chord positions (blue dots). The horizontal row of blue dots
intersects the jet-1,2 merge plane.}
\end{figure}

Two or three adjacent guns (at the vertices of the hexagonal mounting pattern)
are used in the two/three-jet experiments (results of Sec.~\ref{sec:results1}),
as depicted in Fig.~\ref{fig:cartoon}.  Different gas species (N, Ar, Kr, Xe) are used for the two/three-jet experiments.
A side-on perspective of three jets merging with diagnostic positions is shown in Fig.~\ref{fig:diagnostics}.
Diagnostic chords are approximately 70$^\circ$ relative to the plane of the image.
Note that only five chords of the twelve-chord interferometer are used.

\subsection{Setup for merging of six and seven plasma jets}

\begin{figure}[tb]
\includegraphics[width=3.2truein]{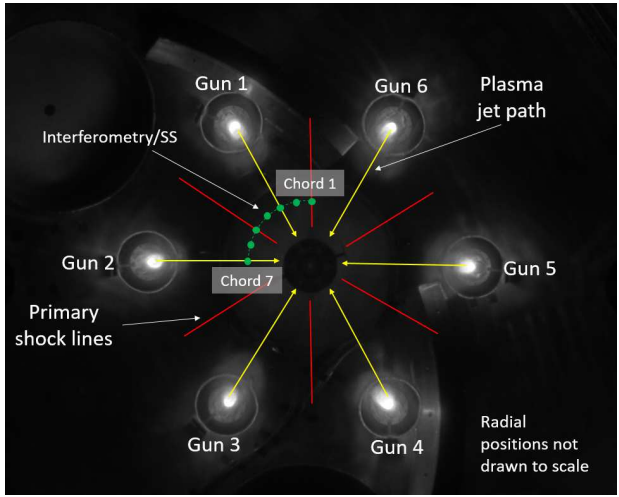}
\caption{\label{fig:6gundiag2} Setup for results of Sec~\ref{sec:results2}: 
annotated, end-on visible image (10-ns exposure), showing the 
inside of the 9-ft.-spherical
vacuum chamber, the location of six of the plasma guns, and locations of seven interferometer and
survey-spectroscopy viewing chords.
The angle between adjacent guns is 23.2$^\circ$. A seventh gun (not shown here) is installed at the center of the 
hexagon.}
\end{figure}

The gun positions and
diagnostic setup for the six/seven-jet experiments (results of Sec.~\ref{sec:results2})
are shown in Fig.~\ref{fig:6gundiag2}.  All the six/seven-jet results shown are for Ar jets.
Interferometer chords (green dots),
normal to the image in Fig.~\ref{fig:6gundiag2}, intersect the jet-propagation axes
14.7~cm from chamber center (5.7~cm in the plane of projection).  Note that only seven chords of the twelve-chord
interferometer are used.

\section{Experimental results:  merging of two and three plasma jets}
\label{sec:results1}

\begin{figure}[!tb]
\includegraphics[width=3.3truein]{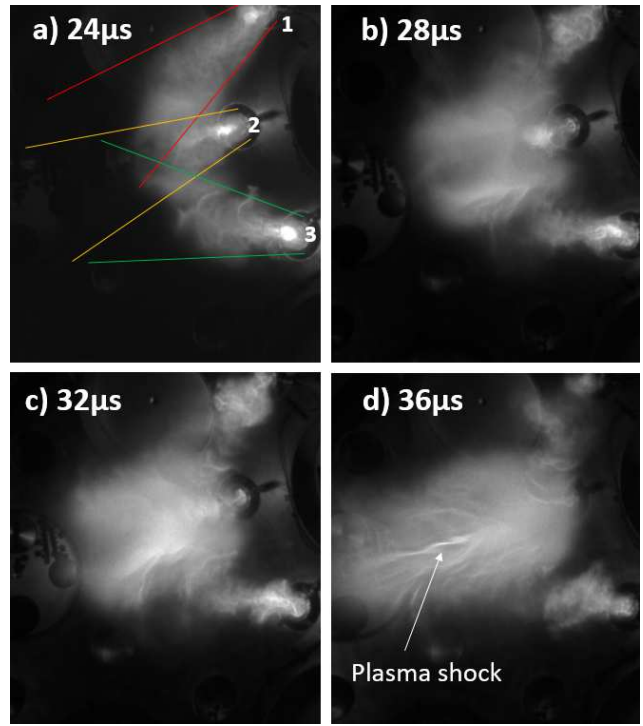}
\caption{\label{fig:3gunimage1} Visible self-emission images (10-ns exposure) of three-jet merging.  (a)~Approximate jet-propagation envelopes.  (b) and (c)~Jets 1,2 and 2,3 have merged.  (d)~All three jets have merged with a narrow
region of intense emission, which we show in this paper to be consistent with a plasma shock.}
\end{figure}

The primary new results of this section are the detailed characterizations of the dynamics and shock formation/evolution
of two- and three-
jet merging (see Fig.~\ref{fig:3gunimage1}), substantially augmenting our recent studies that were focused on ion heating for
two-jet merging,\cite{langendorf2018experimental,langendorf2019experimental} which are based on some of the
same experimental campaigns presented here.  Firstly, interferometer and spectrometer data are
presented to provide space- and time-resolved
$n_e$, $T_e$, and $\bar{Z}$.  Secondly, we discuss
and provide an explanation for the observed shock morphology, including quantitative arguments
based on the collisionality of the jet merging.  All results from Sec.~\ref{sec:results1} are for the setups
shown in Figs.~\ref{fig:cartoon} and \ref{fig:diagnostics}.

\subsection{Inferring plasma parameters and density jumps upon jet merging}
\label{sec:III.A}

Measurements of $n_e$ are obtained via the five interferometry chords shown in Fig.~\ref{fig:diagnostics} (green dots).
The time-resolved, line-integrated $n_e$ averaged over several shots using Ar jets is compared to synthetic data from SFPMax 
in Fig.~\ref{fig:gunsint2} (see Fig.~\ref{fig:NKrXeint} in Appendix~\ref{sec:more_data} for N, Kr, and Xe data).
Some degree of symmetry is expected around chord 10 for all gas species.  The colors of each trace represent a specific laser chord, with similar colors expected to have similar values, i.e., light blue and dark blue.  The higher line-integrated $n_e$ for a particular chord,
e.g., chord 10 in Fig.~\ref{fig:gunsint2}, reflects an increase in $n_e$ due to the formation of a plasma shock as the 
plasma jets merge.  This is consistent with previous, detailed oblique plasma-shock
studies.\cite{merritt2013experimental,merritt2014experimental}  The lack of symmetry about chord 10, which is along the
midplane of jet 1,3 merging as shown in Fig.~\ref{fig:diagnostics}, is due predominantly
to $>20$\% mass imbalance in the different plasma jets due to variations in the GV injection.  The variation is
reduced to $<2$\% for the results in Sec.~\ref{sec:results2}.  The ability to match simulation results to the data was difficult to achieve, requiring customized simulation initial conditions for each jet in an attempt to match the observed asymmetries.  Using the initial, experimentally inferred values of jet density, length, temperature, and axial density profile, we still had to make assumptions about free simulation parameters in the velocity gradients, divergence of the jets (although constrained somewhat by iCCD images), and timing jitter.

\begin{figure}[tb]
\includegraphics[width=3.2truein]{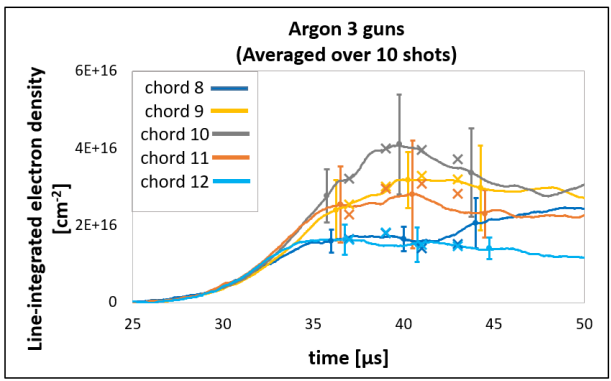}
\caption{\label{fig:gunsint2} Interferometer signals versus time for the chord positions shown in Fig.~\ref{fig:diagnostics}
(green dots) for Ar jets.
Error bars for select times are one standard deviation of the shot-to-shot variation.  Synthetic data from SPFMax simulations
(see Sec.~\ref{sec:mach}) for line-integrated $n_e$ are represented by the x's.  We expect symmetry around chord 10, with similar signals expected for chords 9 and 11 (yellow and orange) and chords 8 and 12 (dark and light blue).}
\end{figure}

Individual shots from Fig.~\ref{fig:gunsint2} (and Fig.~\ref{fig:NKrXeint} of Appendix~\ref{sec:more_data})
are analyzed to better infer the density jump due to the shock structure.  For each shot, ratios are taken between chords with
the highest and lowest line-integrated densities, with the former assumed to represent post-shock and the latter pre-shock 
plasma regions.  The 
high-density shock structure is observed to move across interferometry chords in time as observed in the iCCD images.  
For each species, the maximum ratio between the chords with the highest and lowest line integrated density, 
and their standard deviation, are also determined.  The information is summarized in Table~\ref{tab:densityjump}.  

\begin{table}[b]
\caption{\label{tab:densityjump} Ratio between chords with the highest and lowest line-integrated densities,
as a measure of the shock $n_e$ jump for three-jet merging, based on
the individual shots used in Figs.~\ref{fig:gunsint2} and \ref{fig:NKrXeint}.}
\begin{tabular}{lcccc}
\hline\hline
Gas Species&N&Ar&Kr&Xe\\
\hline
Average&2.9&4.2&6.1&6.6 \\
Standard Dev. &0.8&1.1&1.4&3.1 \\
Maximum&4.6&6.4&8.9&13.6 \\
 \hline
\end{tabular}
\end{table}

The lengths of the merged jets are estimated by taking the full-width half-maximum of the line-integrated $n_e$ traces. 
The merged jet lengths are estimated to be 148, 87, 88, and 57~cm for
N, Ar, Kr, and Xe, respectively.  The long lengths compared to that of individual jets ($<20$~cm) 
are primarily due to speed differences between jets leading to an extended merged-plasma region.  Further
reductions in individual jet lengths and mass/speed variations across jets are needed;
improvements are presented in Appendix~\ref{sec:jets}.

\begin{figure}[tb]
\includegraphics[width=3.2truein]{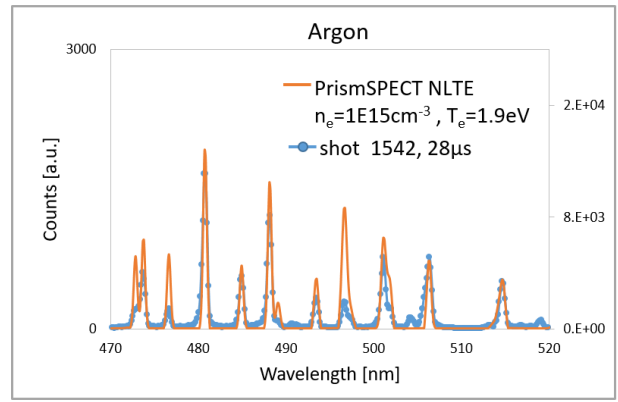}
\caption{\label{fig:nitargspec2} Comparison of experimental and calculated atomic spectra for three-jet
merging for Ar, at the location of chord~10 (the middle
green dot in Fig.~\ref{fig:diagnostics}).  Inferred $n_e$ and $T_e$ are given in the legend.}
\end{figure}

To infer $T_e$ and $\bar{Z}$,
survey-spectroscopy data is compared to non-local-thermodynamic-equilibrium (non-LTE)
calculations of atomic spectra using PrismSPECT, \cite{macfarlane2004ifsa} by noting when specific line transitions
appear or disappear as a function of $n_e$ and $T_e$.
Values of $n_e$ consistent with interferometry measurements are used in PrismSPECT calculations to bound $T_e$
and $\bar{Z}$.  Figure~\ref{fig:nitargspec2} shows the comparison of experimental spectra and PrismSPECT calculations
for three-jet merging using Ar (see Fig.~\ref{fig:NKrXespec} in Appendix~\ref{sec:more_data} for N, Kr, and Xe data).  The 
inferred $T_e \approx 1.2$--3.2~eV (across all four species) including error bars, 
which are given in Table~\ref{tab:tempjump}, for
$n_e=10^{15}$~cm$^{-3}$ (this is the estimated density of the post-merged plasma).  The PrismSPECT calculations indicate 
singly and doubly ionized states with $\bar{Z}=0.9$--1.9 (across all four species).
Interestingly, post-merged plasma jets do not show much spatial nor temporal variation in $T_e$.  There are discrepancies 
between the experimental and calculated line ratios in Figs.~\ref{fig:nitargspec2} and \ref{fig:NKrXespec}, which may be due to 
the line-integrated light collection spanning regions of different plasma parameters. 

\begin{figure}[tb]
\includegraphics[width=3.2truein]{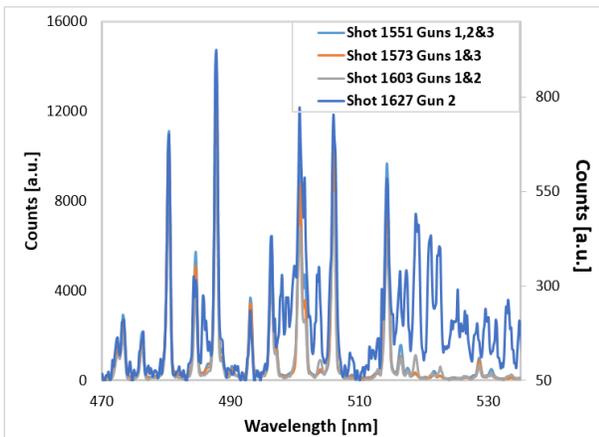}
\caption{\label{fig:SpecGunsComp2} Measured atomic spectra (chord 10, $t=36$~$\mu$s)
for three different argon-jet-merging configurations (shots 1551, 1573, 1603; very similar spectra) and a single jet (shot 1627; very different compared to merging configurations).
The values of pre- and post-merge $T_e$ and $\bar{Z}$ are given in Table~\ref{tab:tempjump}.}
\end{figure}

Figure~\ref{fig:SpecGunsComp2} shows comparisons of experimental and calculated spectra among different 
jet-merging configurations and a single jet for Ar.  Across all merging configurations, the spectra do not significantly change, implying 
similar values of $T_e$ and $\bar{Z}$, but the spectra for a single jet is different.  Comparing with PrismSPECT calculations, the single-jet spectra indicates 
$T_e=1.5$~eV and $\bar{Z}$=0.9 for a density of $5\times 10^{14}$~cm$^{-3}$ (pre-merged density estimate), while the merged-jet spectra indicates
$T_e=1.9$~eV and $\bar{Z}$=1.0 for a density of $1\times 10^{15}$~cm$^{-3}$.  Similar increases in $n_e$, $T_e$, and 
$\bar{Z}$ for merging versus a single jet are observed for all gas species and summarized in Table~\ref{tab:tempjump}.
The increase in
$T_e$ and 
$\bar{Z}$ in merging configurations compared to a single jet was observed in prior experiments,\cite{moser2015experimental} 
which was attributed to frictional heating of electrons from slowing of the oppositely directed ions.  The rate of slowing can be 
estimated using the ion-electron slowing-down rate.  Using the estimated parameters for each species, the slowing-down rate for 
nitrogen is calculated to be significantly larger (factors of approximately 2--3) than the other three species.
If this is the leading 
mechanism for increased $T_e$, then it could explain the larger increase in $T_e$ for nitrogen. 

\begin{table}
\caption{\label{tab:tempjump}}  Experimentally inferred values of $T_e$ and $\bar{Z}$ for single and merged jets, for
four gas species.  The merged values are representative of both two- and three-jet merging.
\begin{tabular}{lcc}
\hline\hline
& single jet & merged jets\\
& [$T_e$(eV), $\bar{Z}$] & [$T_e$(eV), $\bar{Z}$]\\
\hline
N&[1.3$\pm$0.3, 0.8$\pm$0.2]&[2.8$\pm$0.4, 1.4$\pm$0.4]\\
Ar& [1.5$\pm$0.3, 0.9$\pm$0.3]&[1.9$\pm$0.5, 1.0$\pm$0.1]\\
Kr&[1.2$\pm$0.4, 0.8$\pm$0.4]&[1.9$\pm$0.5, 1.2$\pm$0.3]\\
Xe&[1.4$\pm$0.4, 1.0$\pm$0.3]&[1.7$\pm$0.5, 1.5$\pm$0.4]\\
 \hline
 \end{tabular}
\end{table}

\subsection{Prospect of shockless jet merging}

Earlier two-jet merging experiments using mini-railguns on PLX demonstrated oblique plasma-shock formation in highly 
collisional
regimes.\cite{merritt2013experimental,merritt2014experimental}  Simulations of spherical-plasma-liner
formation via merging
jets in the collisional, hydrodynamic limit showed a ``cascade of shocks'' as jets merged and during
the subsequent liner implosion.\cite{kim2013structure}
More recently, collisional plasma-shock formation was reproduced using coaxial plasma guns,\cite{langendorf2018experimental,langendorf2019experimental} but those experiments also showed
that jet--jet interpenetration without
shock formation could occur in less-collisional regimes.\cite{langendorf2018experimental,langendorf2019experimental}  

This leads to the idea of deliberately operating in a less-collisional jet-merging regime to
avoid shock formation, thereby minimizing the seeding of density perturbations in plasma-liner formation as desired for
PJMIF\@.\cite{langendorf2019experimental} 
The three-jet-merging results of this paper point to further constraints.  We will show that in these experiments,
despite the likelihood
that the merging of only jets 1,2 and 2,3 and 1,3 are occurring in an interpenetrating regime, the collisionality is increased
to form a shock when all three jets merge.

For these experiments, interferometer data do not show large density jumps associated with a shock front
passing through for any of the two-jet configurations shown in Fig.~\ref{fig:cartoon}.  This is supported by 
two-jet interferometer data discussed later in Figs.~\ref{fig:machint}(b) and (c), where maximum density jumps are $<2$.  
Furthermore, the iCCD images in Figs.~\ref{fig:3gunimage1}(b) and (c)
show a diffuse bright region at the merge regions between jets 1,2 and 2,3.  The diffuse emission morphology 
was shown previously to correspond to an interpenetrating, shockless jet-merging situation.\cite{moser2015experimental,langendorf2018experimental,langendorf2019experimental}

However, despite the two-jet merging leading to shockless, semi-collisional interpenetration, the merging of all
three jets leads to a shock, as seen in Fig.~\ref{fig:3gunimage1}(d) and the $>2$ density jump in Fig.~\ref{fig:machint}(a).  Here,
we show that the presence of a shock in three-jet merging, despite the absence of shocks in two-jet merging,
is due to dynamically increasing $T_e$ and $\bar{Z}$ 
due to jet merging.  This is exacerbated by the presence of impurity titanium, which leads to even higher $\bar{Z}$.  We denote
this the ``Moser effect,'' which was first observed in head-on jet-merging experiments on PLX;\cite{moser2015experimental} a 
similar effect has been recently observed in laser-driven experiments.\cite{young19pop}

\begin{figure}[!tb]
\includegraphics[width=3.34truein]{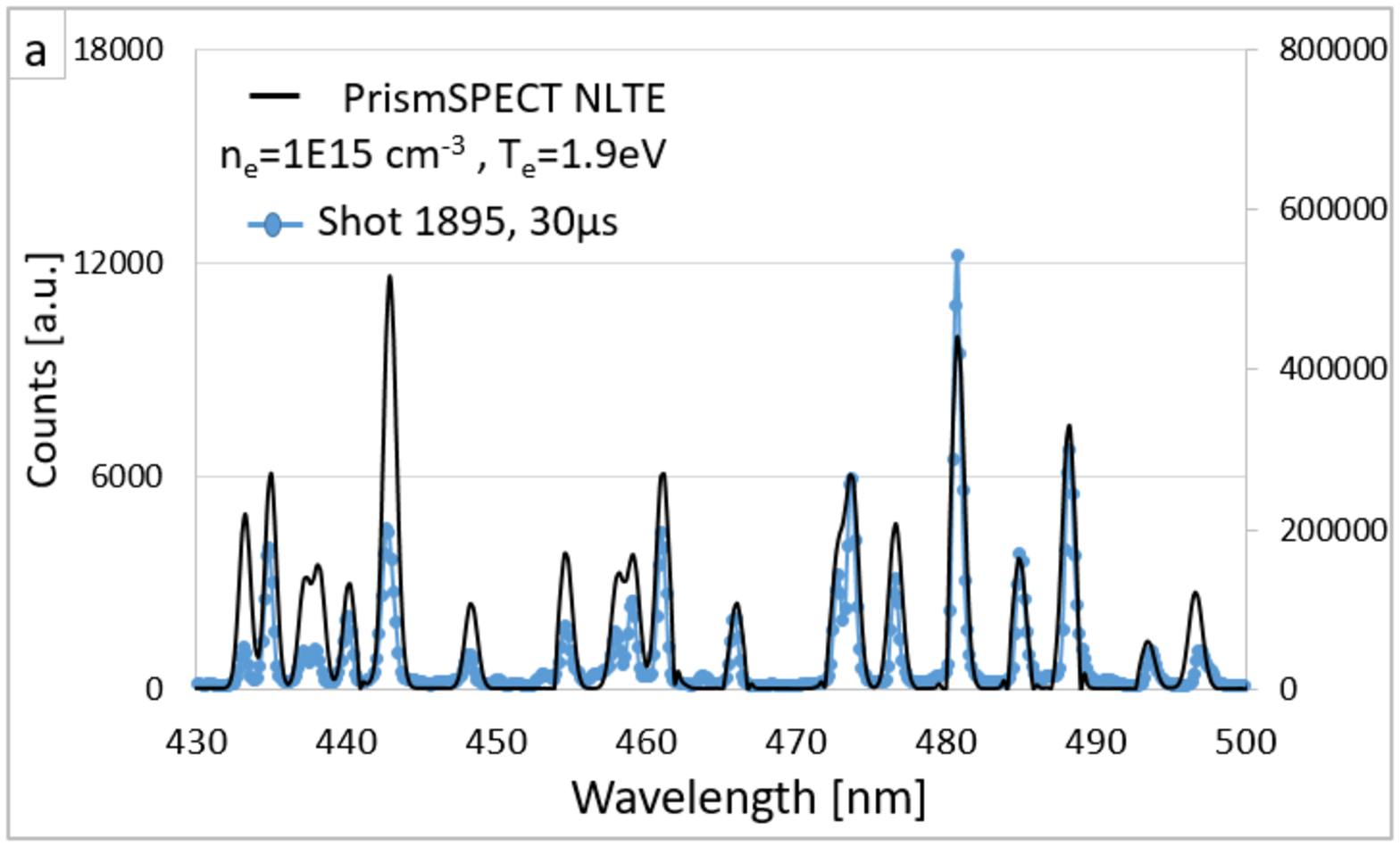}
\includegraphics[width=3.34truein]{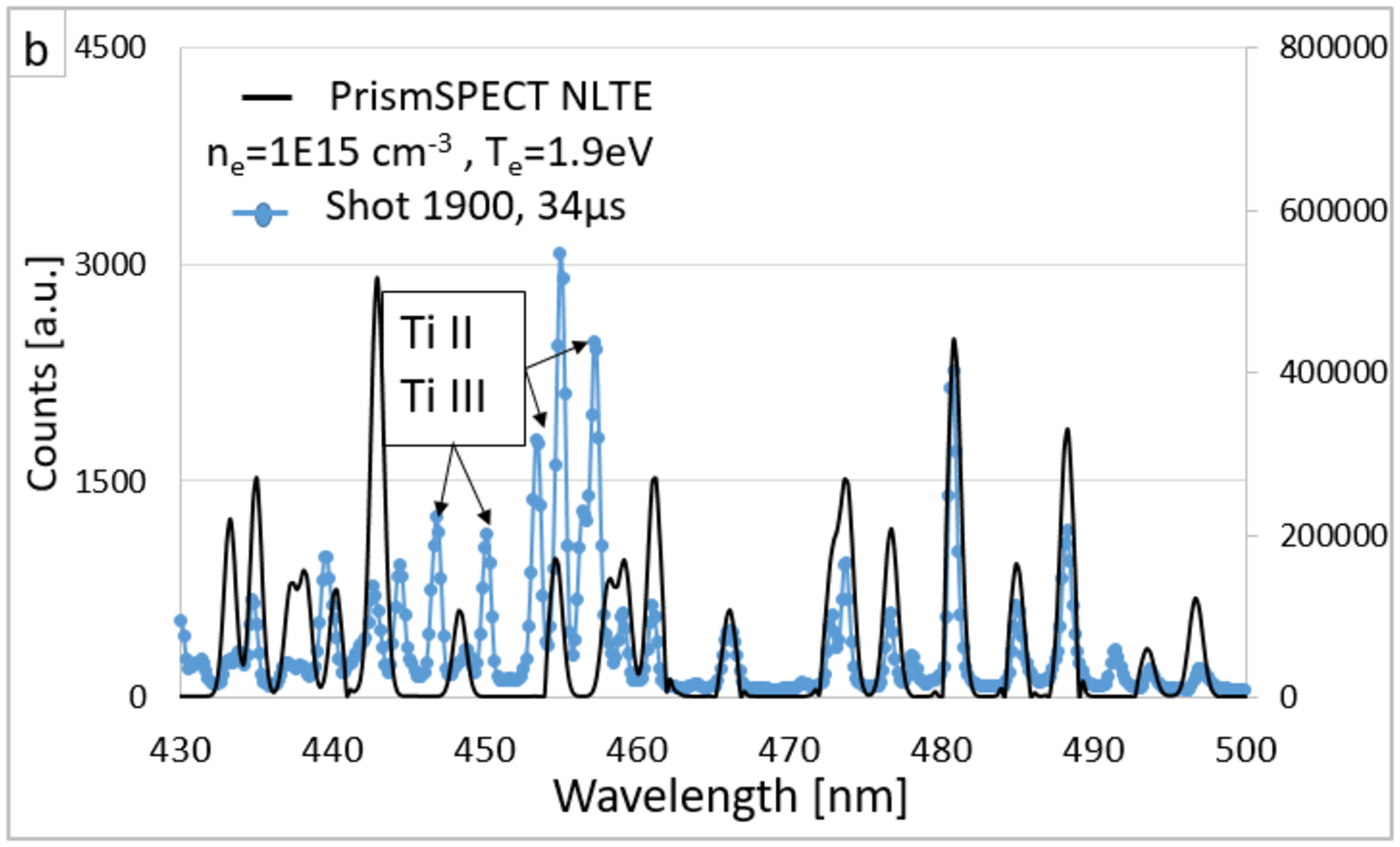}
\caption{\label{fig:titanium}  Survey-spectroscopy data (chord-10 position) for three-jet merging (Ar) and fits to PrismSPECT calculated spectra (no Ti), for (a) leading front of the jet (no Ti) and (b) bulk of the jet (with Ti).}
\end{figure}

The presence of titanium, from the inner electrode of the coaxial plasma guns, is evident from side-on
survey-spectroscopy data (Fig.~\ref{fig:titanium}) taken at the chord-10 position (see Fig.~\ref{fig:diagnostics}).
To estimate the fraction of impurities in the plasma jets, vacuum-chamber pressure rise for gas injection only is
compared against that with plasma-jet discharges, which presumably ablates some electrode material.
This comparison suggests that the plasma jet has 80\% injected gas species (in this case Ar) and 20\% impurities.  The 
three Ar jets arrive at the chord-10 position at about 27~$\mu$s (see Fig.~\ref{fig:gunsint2}). The titanium impurities, 
Ti~{\sc ii} and Ti~{\sc iii}, appear around 34~$\mu$s, with about 10~$\mu$s of leading edge of the jet having little to no measurable impurities.  The best-fit PrismSPECT calculations indicate $\bar{Z}=1.0$ for
the data at $t=30$~$\mu$s and $\bar{Z}=2.0$ for $t=34$~$\mu$s.
We estimate that only $\sim5$\% of the jet mass is in the leading edge where there is little-to-no titanium.  Unfortunately the spectroscopy analysis of impurities was only conducted for argon jets.

Combining all the information on plasma parameters that we have gathered here and elsewhere,\cite{langendorf2019experimental} we provide quantitative estimates of the collisionality for
three-jet merging for both cases of with and hypothetically without impurities.  As 
discussed previously,\cite{merritt2014experimental}
the ion--ion interpenetration length is 
\begin {equation}
L_{ii,s} = \sum_{i^\prime}\frac{v}{4\nu_{ii^\prime,s}},
\label{eq:Lii}
\end {equation}
where $\nu_{ii^\prime,s}$ is the ion--ion slowing frequency in the fast limit ($\gg \nu_{ie,s}$ for the present parameters),\cite{huba2018nrl}
\begin {equation}
\label{eq:nu}
\nu_{ii^\prime,s}=9\times 10^{-8} n_i^\prime \Lambda_{ii^\prime}\bar{Z^\prime}^2\bar{Z}^2\left(\frac{1}{\mu}+\frac{1}{\mu^{\prime}}\right)\frac{\mu^{1/2}}{\epsilon^{3/2}},
\end {equation}
and $v = 2 v_{jet} \sin\theta$ is the relative normal speed between counter-streaming plasma jets, $\theta$ is the half
angle between jets, $n_{i}$ is the ion density, $\Lambda_{ii^\prime}$ is the Coulomb logarithm for counter-streaming
ions in the presence of warm electrons,\cite{merritt2014experimental} $\mu$ is the ion/proton mass ratio, and $\epsilon$ (eV) is the energy associated with $v$.  In Eq.~(\ref{eq:Lii}), the factor of 4  is the integral effect of the slowing
down,\cite{messer2013nonlinear} and the summation is over all gas species.  Prime and unprimed variables refer to field and test particles, respectively.\cite{huba2018nrl}
Table~\ref{tab:collisionality} shows calculated $L_{ii,s}$ for each jet species using
averaged plasma parameters.  The impurity (Ti) speed is assumed to be the same as for the majority species.  
The $T_e$ and $\bar{Z}$ are determined from survey spectroscopy for post-merging of jets, with the same values assumed for 
Ti across all jet species, while $n_e$ is from interferometry.  The half-angle $\theta=20.5^{\circ}$ between jets 1,3
is used in these estimates.  Two estimations of $L_{ii,s}$ are calculated for the 80\%/20\% mixtures, i.e.,
with the majority and impurity species each treated as the test particle, respectively.
For both cases, $L_{ii,s}$ values are summed over both field ion species, as indicated in Eq.~(\ref{eq:Lii}).  
To be in a highly collisional limit with shock formation, $L_{ii,s}$ should be
$\ll 30$~cm (jet system size, from the iCCD images).   Our calculations show that using
the parameters from Table~\ref{tab:tempjump} (for jets with no impurities) would result in substantial
interpenetration.  This is consistent with the lack of shock structure observed in the iCCD images and the interferometry chords 
for the merging of jets 1,3.  However, as seen in Table~\ref{tab:collisionality}, the merging of all three jets leads to an
increase in $T_e$ and $\bar{Z}$ compared to Table~\ref{tab:tempjump}.  This,
along with the presence of Ti impurities with even higher $\bar{Z}$, leads to an increase in collisionality and
reduction in $L_{ii,s}$ and therefore the
formation of a shock along the center line of the three jets, as observed in Fig.~\ref{fig:3gunimage1}(d).

\begin{table*}[tb]
\caption{\label{tab:collisionality} Collisionality of jet merging is evaluated based on Eqs.~(\ref{eq:Lii}) and (\ref{eq:nu}),
using $\theta=20.5^\circ$ corresponding to the half-angle between jets 1,3.
Results are shown assuming 100\% majority-species jets (top half of table) and an 80\%/20\% mixture of majority-species/Ti
jets (bottom half of table).  The two values of $L_{ii,s}$ shown for the 80\%/20\% mixtures are for majority and Ti ions
as the test particles, respectively.}
\begin{tabular}{lccccc} 

\hline\hline
Species&N&Ar&Kr&Xe\\
\hline\hline
$v$ (km/s)&36.5&29.4&39.8&19.2\\
\hline
$n_{e}$ ($10^{14}$$cm^{-3}$)&3.0$\pm$0.3&2.9$\pm$0.4&3.3$\pm$2.0&2.6$\pm$0.3\\

\hline
$T_{e}$(eV)&2.8$\pm$0.4&1.9$\pm$0.4&1.9$\pm$0.5&1.7$\pm$0.5\\
\hline
$\bar{Z}$&1.4$\pm$0.4&1.0$\pm$0.1&1.2$\pm$0.3&1.5$\pm$0.4\\
\hline
$L_{ii,s}$(cm)&1.5&18.0&109.0&8.0\\
\hline\hline
Species&80$\%$N/20$\%$Ti&80$\%$Ar/20$\%$Ti&80$\%$Kr/20$\%$Ti&80$\%$Xe/20$\%$Ti\\
\hline\hline
$v$ (km/s)&36.5&29.4&39.8&19.2\\
\hline
$n_{e}$ ($10^{14}$$cm^{-3}$)&2.4/0.6$\pm$0.2/0.1&2.4/0.5$\pm$0.3/0.1&2.6/0.7$\pm$1.6/0.4&2.1/0.5$\pm$0.2/0.1\\

\hline
$T_{e}$(eV)&2.8/1.4$\pm$0.4/0.2&1.9/1.4 $\pm$0.4/0.2 &1.9/1.4 $\pm$0.5/0.2&1.7/1.4$\pm$0.5/0.2 \\
\hline
$\bar{Z}$&1.4/2.0$\pm$0.4/0.4&1.9/2.0$\pm$0.1/0.4&1.2/2.0$\pm$0.3/0.4&1.5/2.0$\pm$0.4/0.4\\
\hline
$L_{ii^\prime,s}$(cm)&1.3/3.0&8.8/5.0&50.0/2.6&4.7/0.4\\
 \hline
 \end{tabular}
\end{table*}

To summarize the primary points of this subsection, it is possible to have semi-collisional, interpenetrating, shockless
jet merging, which may be desirable for plasma-liner formation to minimize the amplitude of density non-uniformities.
However, jet merging leads to slightly elevated $T_e$ and $\bar{Z}$ that is further exacerbated by the presence
of jet impurities, like Ti, that tend to have even higher $\bar{Z}$.  Because collisionality $\sim \bar{Z}^4$, it is easy to
transition into a regime with shock formation, i.e., ``Moser effect,'' for multiple-jet merging in the presence of impurities.
This points to a need to further reduce the impurity content in the plasma jets 
to enable shockless jet merging as an option for plasma-liner formation.

\subsection{Liner Mach-number degradation}
\label{sec:mach}

Maintaining a high liner Mach number $M$ to enable efficient target compression has been identified as important for
the viability of PJMIF\@.\cite{parks08pop,langendorf2017semi}  The recent, related results on 
shock ion heating in two-jet merging experiments showed that ion heating to tens of eV occurs upon jet merging for cases with and without shock formation.\cite{langendorf2019experimental}  This degrades $M$, but due to very fast cooling of
the ions on the few-eV electrons over a several-$\mu$s timescale, $M$ should quickly increase.
In this work, we estimate $M$ and its evolution using the previously reported high-resolution Doppler-spectroscopy 
data (providing ion temperature $T_i$)\cite{langendorf2018experimental,langendorf2019experimental} and the 
interferometer/survey-spectroscopy data of this paper.  We use the data to benchmark hydrodynamic simulations
of jet-merging using the three-dimensional SPFMax smooth-particle hydrodynamics (SPH) code.\cite{schillo2019suite}
Simulations of fully spherical plasma-liner formation and implosion,
benchmarked to our localized jet-merging experimental data, show that the liner-averaged $M$
remains high even with the transient localized reduction due to shock heating.

\begin{figure}
\includegraphics[width=3.2truein]{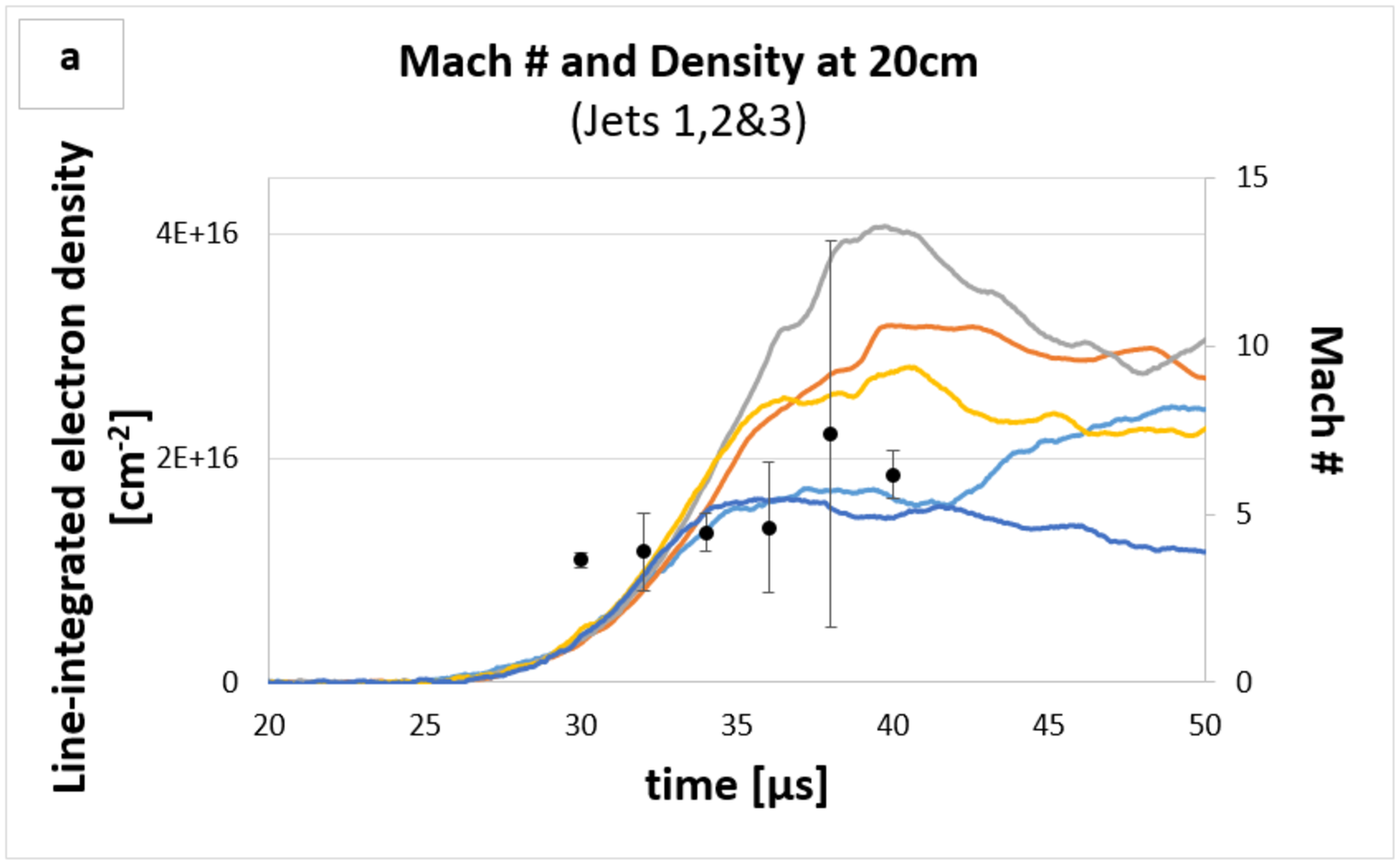}
\includegraphics[width=3.2truein]{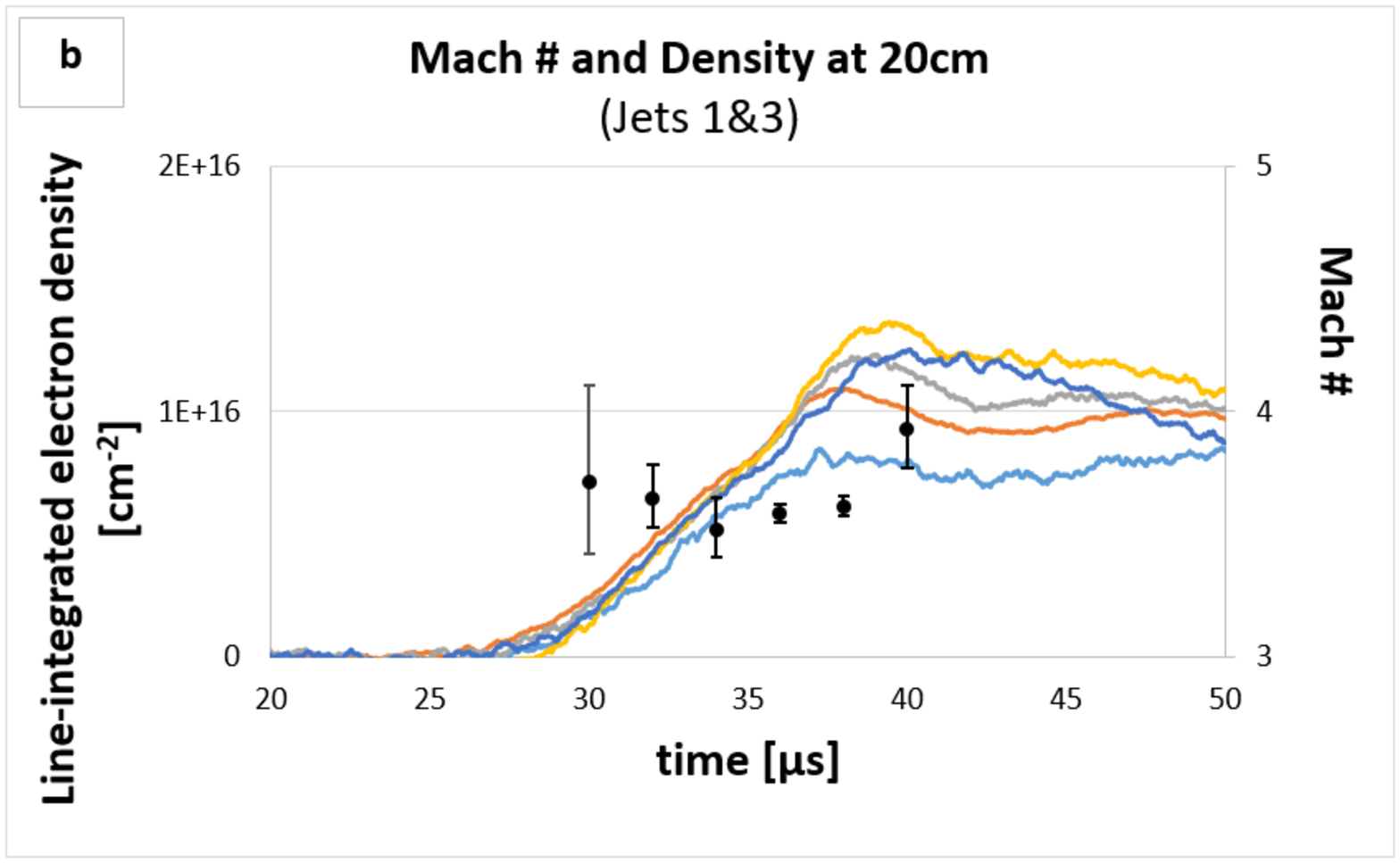}
\includegraphics[width=3.2truein]{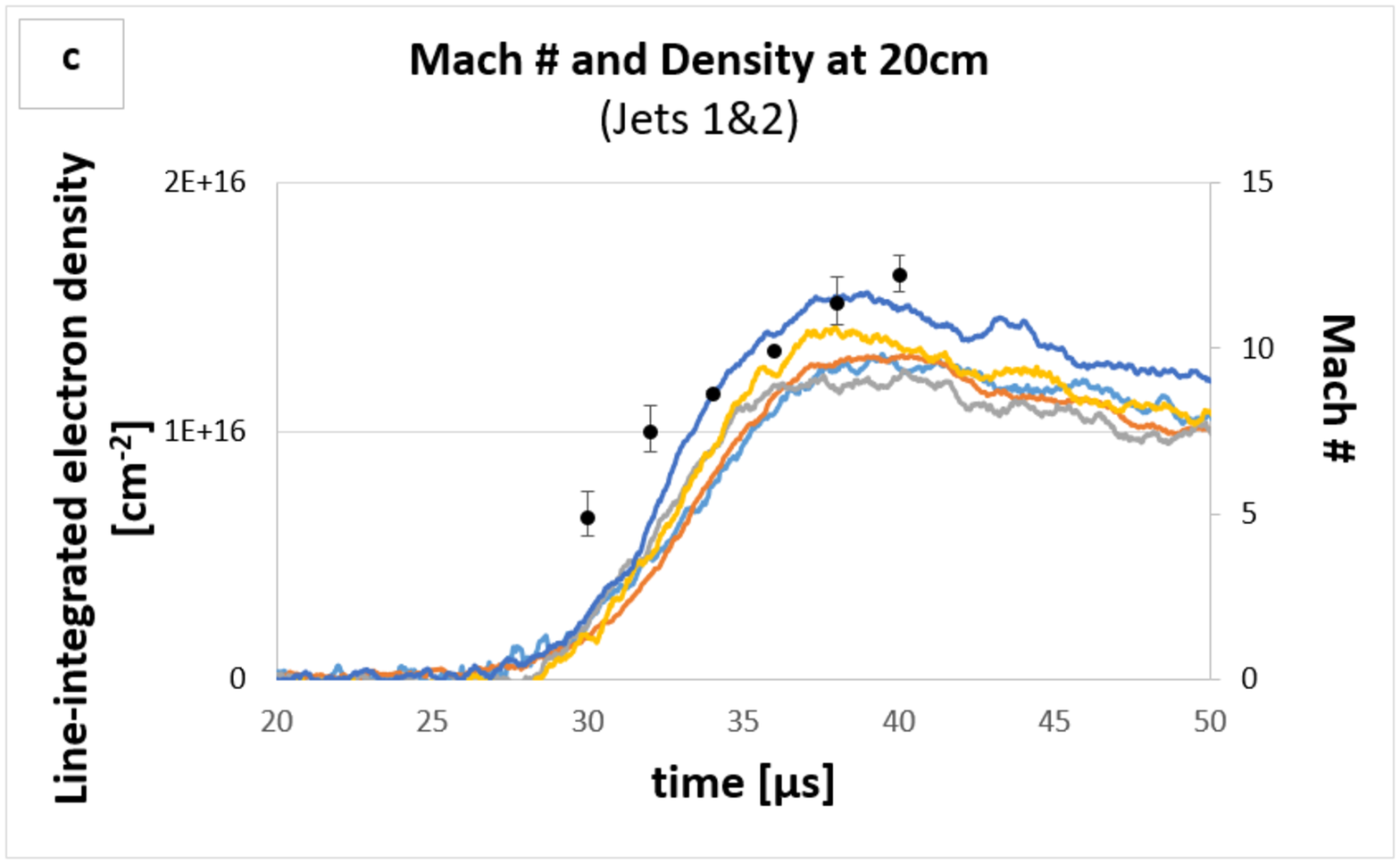}
\caption{\label{fig:machint}Line-integrated $n_e$ (colored lines) and experimentally inferred values of $M$ (black
dots) for merging of jets
(a) 1,2,3, (b) 1,3, and (c) 1,2.  The data are all taken at 20~cm from chamber center, at the jet-2
propagation axis for (a) and (b) and the jet 1,2 midplane for (c) (see Fig.~\ref{fig:diagnostics} for positions).  
Colors correspond to the chord numbers shown in the legends of Fig.~\ref{fig:gunsint2}.}
\end{figure}

Doppler-spectroscopy data, providing time- and space-resolved
$T_i$,\cite{langendorf2018experimental,langendorf2019experimental} was obtained at the positions of the blue dots in 
Fig.~\ref{fig:diagnostics}.
Along with $T_e$ and $\bar{Z}$ data presented in this paper, the time- and space-resolved 
$M=v_{\rm jet}/C_s$ is calculated in the jet-merging and shock regions,
where $C_s=(\gamma p/\rho)^{1/2}= [\gamma (\bar{Z}T_e + T_i)/m_i]^{1/2}$.
We use the gas sound speed
definition due to the very high collisionality, which invalidates the isothermal assumption for electrons.
Adiabatic index $\gamma = 5/3$ is assumed due to the unconstrained three-dimensional motion of both electrons and ions
in our unmagnetized plasmas (note:  $\gamma$ can be $<5/3$ for low charge states of
Ar, Kr, Xe,\cite{hsu2018physics} so we may be underestimating $M$).  
Figure~\ref{fig:machint} shows side-on interferometry and inferred $M$ using Doppler-spectroscopy measurements of 
$T_i$ at approximately
20~cm from chamber center (see Fig.~\ref{fig:diagnostics}) for three different argon jet-merging configurations.
The interferometry traces shown in Fig.~\ref{fig:machint}(a) and (b) cross the region where the plasma jets merge and
plasma shocks can form.   Figure~\ref{fig:machint}(a) shows a large ratio ($>2$) of line-integrated $n_e$ between
the chords with the highest and lowest values, consistent with the presence of a shock,
while Fig.~\ref{fig:machint}(b) shows a much smaller ratio ($<2$).  For the three-jet case [Fig.~\ref{fig:machint}(a)], the 
jet merging leads to collisional ion heating, resulting in a low $M$, followed by subsequent cooling of the ions through
collisional equilibration with the cool electrons, which allows $M$ to increase back up to nearly 7.  For the two-jet case at 
larger angle [Fig.~\ref{fig:machint}(b)], ion heating is observed; however, the lower density inhibits radiative cooling
and results in a lower $M$.  For the two-jet case at smaller angle 
[Fig.~\ref{fig:machint}(c)], the collisional ion shock heating leads to a low $M$ followed by cooling that leads to a high Mach 
number.

\begin{figure}
\includegraphics[width=2.8truein]{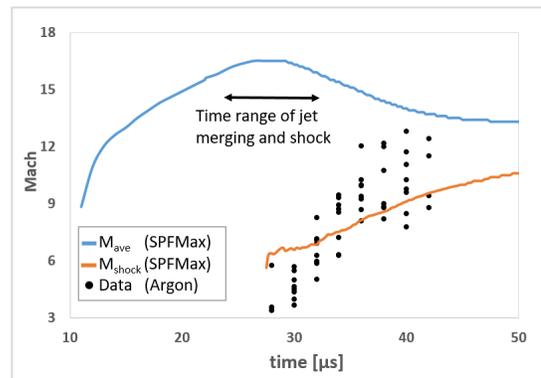}
\caption{\label{fig:Mach_degradation} Mach number versus time:  black circles and orange line are the experimentally inferred 
and simulated values, respectively, within a two-jet merging region (guns 1,2).  Solid blue line is the simulated, liner-averaged
Mach number from the same spherical simulation, showing that the liner-average value remains above $\sim10$ even while the
localized value within the jet-merging region falls below 10.}
\end{figure}

Hydrodynamic simulations using the SPH code SPFMax\cite{schillo2019suite} are conducted and benchmarked against
the two- and three-jet merging data in this paper.  SPFMax and its predecessor codes have been used extensively for both
one- and three-dimensional modeling of imploding plasma liners.\cite{cassibry2012tendency,cassibry2013ideal}
Figure~\ref{fig:Mach_degradation} presents results for two-jet argon merging compared to simulations using SPFMax,
showing (1)~experimentally inferred $M$ values within the jet-merging region of jets 1,2 (black dots);
(2)~simulated local values of $M$ at the region of merging (red line); and (3)~simulated spherical-liner-averaged 
$M$ (blue line), indicating that the liner-averaged $M$ remains $\gtrsim 10$ despite the localized value in the jet-merging 
region falling well below 10.  The experimentally inferred $M$ values use $T_i$ measurements from the
blue dots along the jet 1,2 midplane shown in Fig.~\ref{fig:diagnostics}.
The ability for the global $M$ to remain $>10$ and the localized $M$ to rise back to $\gtrsim 10$ 
is important for the viability of the PJMIF concept because degradation in $M$ would lead to an increase in liner 
spreading, which would degrade the liner's ability to compress a plasma target to fusion-relevant 
conditions.\cite{langendorf2017semi}

\section{Experimental results:  merging of six and seven plasma jets with improved jet-to-jet mass balance}
\label{sec:results2}

The primary objectives of this section are to experimentally characterize the structure of the section of a plasma
liner formed by six and seven hypersonic merging plasma jets
(see Fig.~\ref{fig:6gundiag2}), and to demonstrate improved mirror symmetry in the liner
structure about the jet-propagation axes due to the improved jet-to-jet mass balance compared to
earlier work.\cite{hsu2018experiment}
Experimental evidence for the improved jet-to-jet balance, achieved by upgraded
GVs and fine-tuned series (ballast) impedances of the GVs, is presented in Appendix~\ref{sec:jets}.
Figure~\ref{fig:iccd67gun}
\begin{figure*}
\centering
\includegraphics[width=5.5truein]{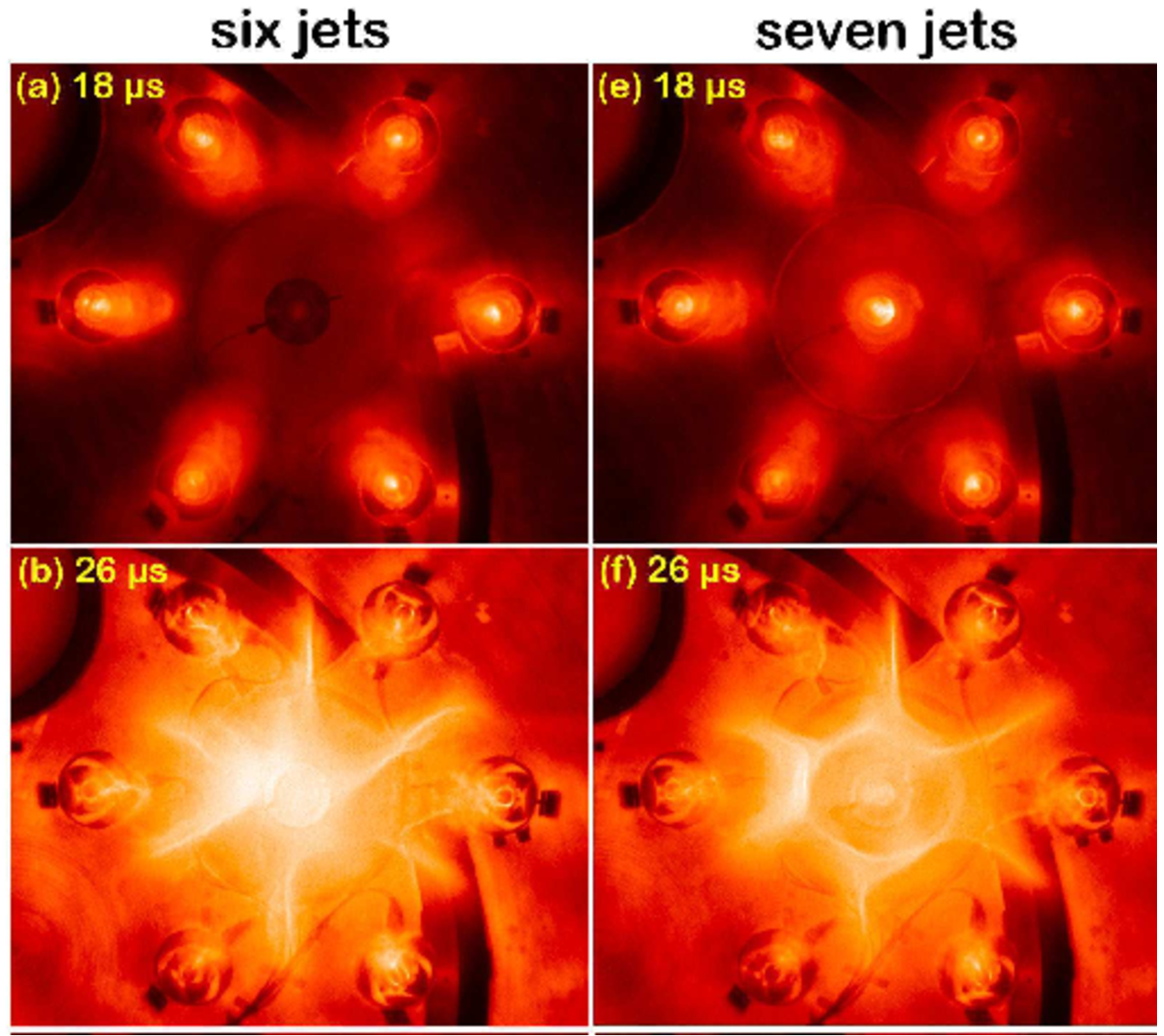}
\includegraphics[width=5.5truein]{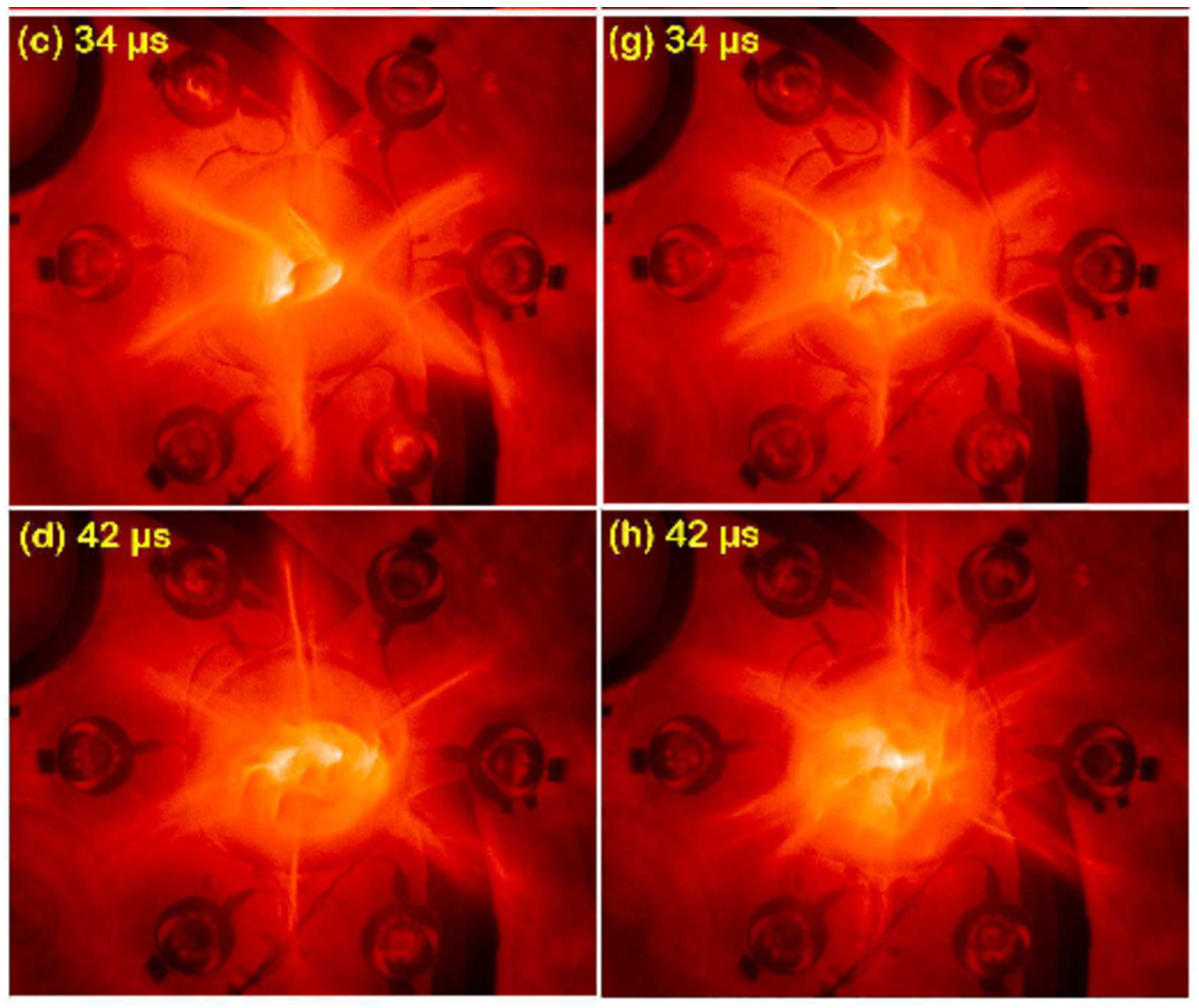}
\caption{\label{fig:iccd67gun}  End-on, self-emission visible images (10-ns exposure, false color) of the merging of
six and seven argon plasma jets.}
\end{figure*}
shows end-on images of the merging of six and seven plasma jets.
Primary shocks (defined here as the oblique formation of plasma shocks and appearing as radial spokes) are observed to form between adjacent 
jets\cite{langendorf2018experimental,langendorf2019experimental},
for both six- and seven-jet cases by $\approx$26~$\mu$s [Figs.~\ref{fig:iccd67gun}(b) 
and (f)].  The seven-gun configuration also forms primary shocks [hexagonal pattern, Fig.~\ref{fig:iccd67gun}(f)] between the six outer jets and the seventh 
jet.  The initially merged plasmas from adjacent jets appear to all merge and form a secondary shock (defined here as the merging of post-shock plasma regions) near the center, pointing
out of the page [Figs.~\ref{fig:iccd67gun}(d) and (h)].  In this section, we present interferometry data to 
characterize time- and space-resolved $n_e$ and $T_e$ associated with the jet merging.
Comparisons are made with
synthetic data from SPFMax\cite{schillo2019suite} three-dimensional hydrodynamic simulations.
Previously published FronTier hydrodynamic simulations
provide information on the influence
of jet-to-jet variations (in both mass and timing) on the structure and quality of plasma-liner formation,\cite{shih2019simulation}
showing good agreement with the images in Fig.~\ref{fig:iccd67gun} for mass variations of $<10$\% and timing
variations of $\leq$ 100~ns.

\subsection{Spatial density uniformity}
 
With reasonably good jet-to-jet balance, it is expected that the density profile along the interferometer chord positions
will exhibit mirror symmetry about chords 3 and 5, i.e., similar values for chords 1 and 5, chords 2, 4, and 6, and chords 3 and 7 (see Fig.~\ref{fig:6gundiag2}).  This was not observed
in earlier six-jet merging experiments due to poor jet-to-jet balance.\cite{hsu2018experiment}  The new data presented
here shows a significant improvement in the mirror symmetry of the density profile about chords 3 and 5,
and a gentler density gradient compared to those in the
synthetic data from from both SPFMax\cite{schillo2019suite} and FronTier.\cite{shih2019simulation}

\begin{figure*}[!tb]
\includegraphics[width=6truein]{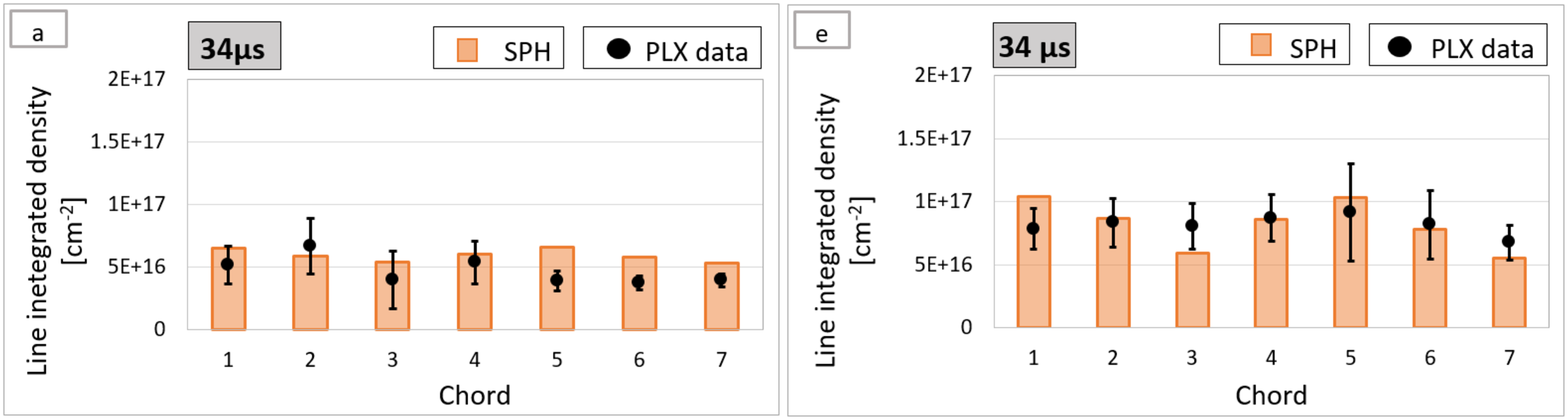}
\includegraphics[width=6truein]{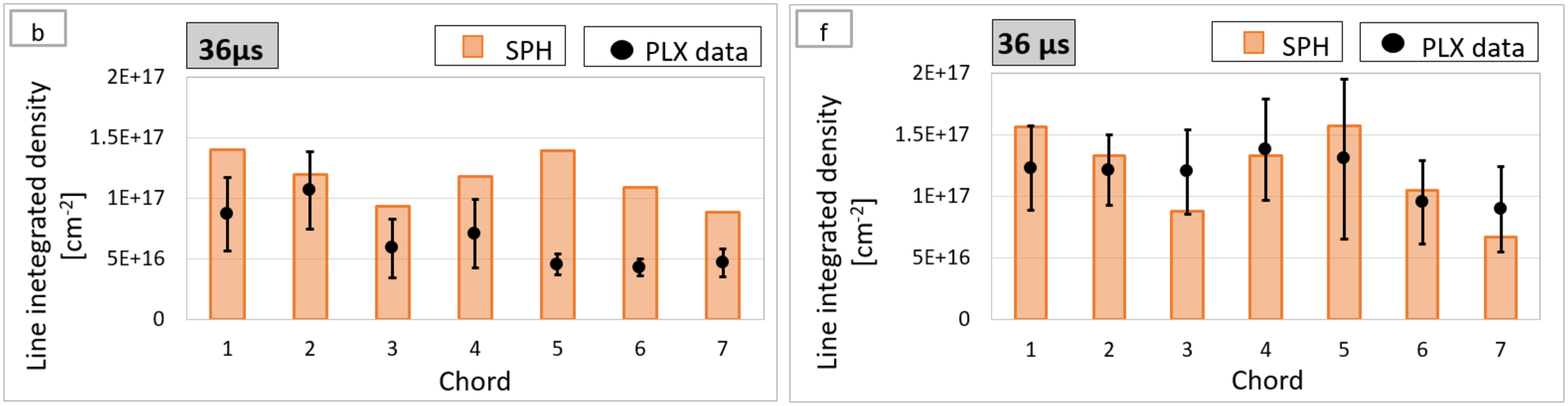}
\includegraphics[width=6truein]{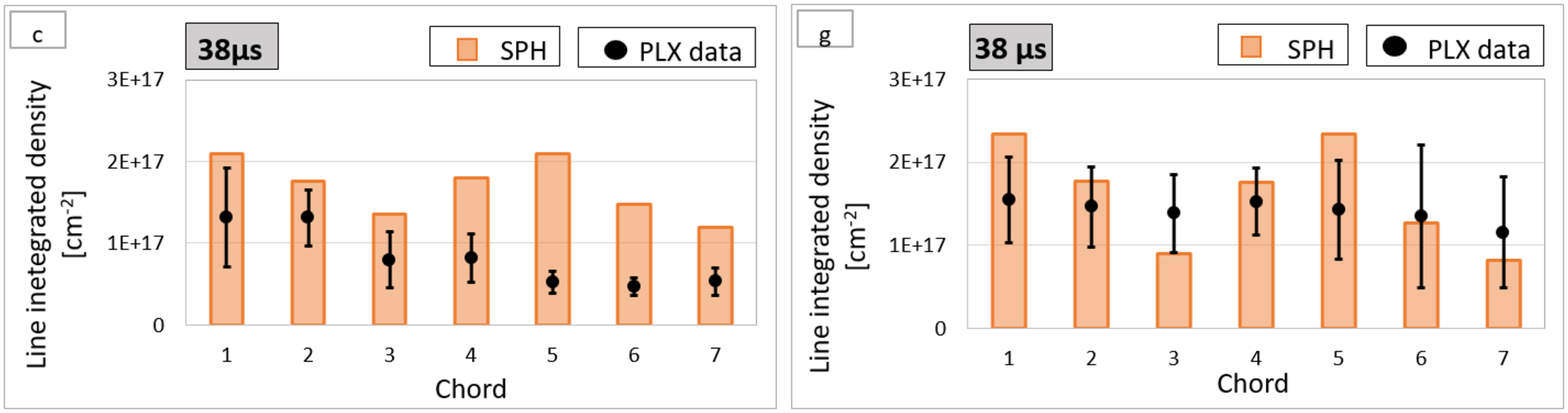}
\includegraphics[width=6truein]{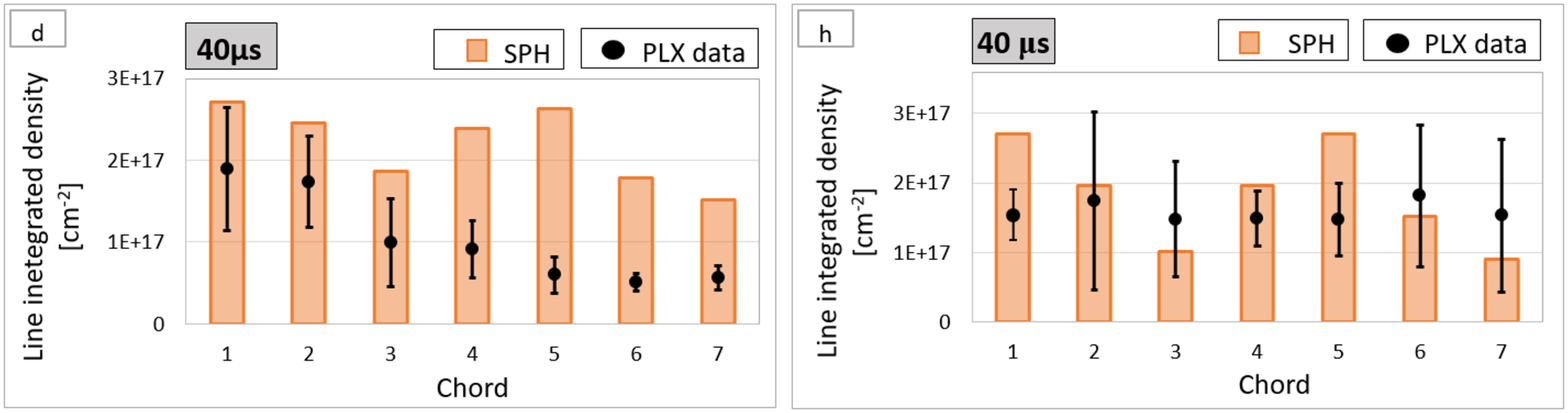}
\caption{\label{fig:6gunint}  Experimental and synthetic SPFMax\cite{schillo2019suite}
data (dots and bars, respectively) for line-integrated $n_e$ from the interferometer chord positions shown in 
Fig.~\ref{fig:6gundiag2}.  Panels (a)--(d) are prior to
and (e)--(h) are after GV/ballast upgrades described in Appendix~\ref{sec:jets}.
Experimental data are averaged over ten shots; the error
bars represent one standard deviation of the shot-to-shot variation.}
\end{figure*}

The seven-chord interferometer provides time- and space-resolved, line-integrated $n_e$ at the positions
shown in Fig.~\ref{fig:6gundiag2} for six-jet merging.  
Figure~\ref{fig:6gunint} shows the interferometry data compared to synthetic data from SFPMax at different times,
both before (left column) and after the GV and ballast upgrades (right column), which are described
in Appendix~\ref{sec:jets}. The perfect mirror symmetry about
chords 3 and 5 is of course seen in the synthetic data, and a substantial improvement is seen in the
experimental data.  The results obtained prior to the GV/ballast upgrade [Fig.~\ref{fig:6gunint}(a)--(d)]
show unbalanced merging of six jets as indicated by the systematically 
low line-integrated $n_e$ measured in chords 5, 6, and 7.  
The results obtained after the upgrade [Fig.~\ref{fig:6gunint}(e)--(h)]
show a more-balanced merging of the plasma jets with improved mirror symmetry about chords 3 and 5.
However, the symmetry is not perfect and is likely due to the timing and velocity jitter discussed earlier.  
Nevertheless, better agreement is now
achieved between experimental and synthetic SPFMax data. 

\begin{figure}[!tb]
\includegraphics[width=3.4truein]{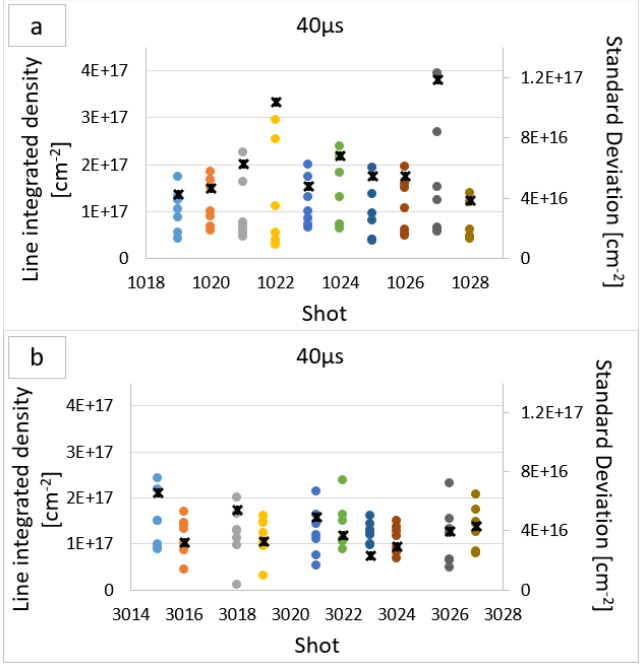}
\caption{\label{fig:shotvar}   Line-integrated $n_e$ ($t=40$~$\mu$s) from all seven interferometer chords 
(colored dots) are plotted
versus individual shot numbers from experiments (a) before and (b) after the GV/ballast upgrades.
Standard deviation across chords is given by the black x's (right-hand y-axis).}
\end{figure}

To further quantify the density non-uniformity for a single shot, individual shots are analyzed from six-jet merging 
experiments before and after the GV/ballast upgrades.  Figure~\ref{fig:shotvar}
shows that the variation across interferometer chords for individual shots is reduced after
the GV/ballast upgrades, with standard deviations across chords falling from
0.4--$1.2\times 10^{17}$~cm$^{-2}$ before the upgrades to 2--$7\times 10^{16}$~cm$^{-2}$ after the upgrades.  While
continued refinements to the jet-to-jet balance will further improve the liner density uniformity, there will be a minimum
level of liner non-uniformity from the jet-merging dynamics, as presented and discussed in Sec.~\ref{sec:results1}.

\subsection{Constancy of electron temperature in space and time}

Knowledge of space- and time-resolved $T_e$ is important for characterizing a section of a spherically 
imploding plasma liner.  It affects the mean-charge state $\bar{Z}$, which has a strong influence on inter-jet collisionality
through a $\bar{Z}^4$ dependence [see Eq.~(\ref{eq:nu})], and it affects the electrical resistivity $\sim T_e^{-3/2}$, 
which affects the eventual magnetic-diffusion dynamics when the liner compresses a magnetized plasma target. 

\begin{figure}
\includegraphics[width=3.4truein]{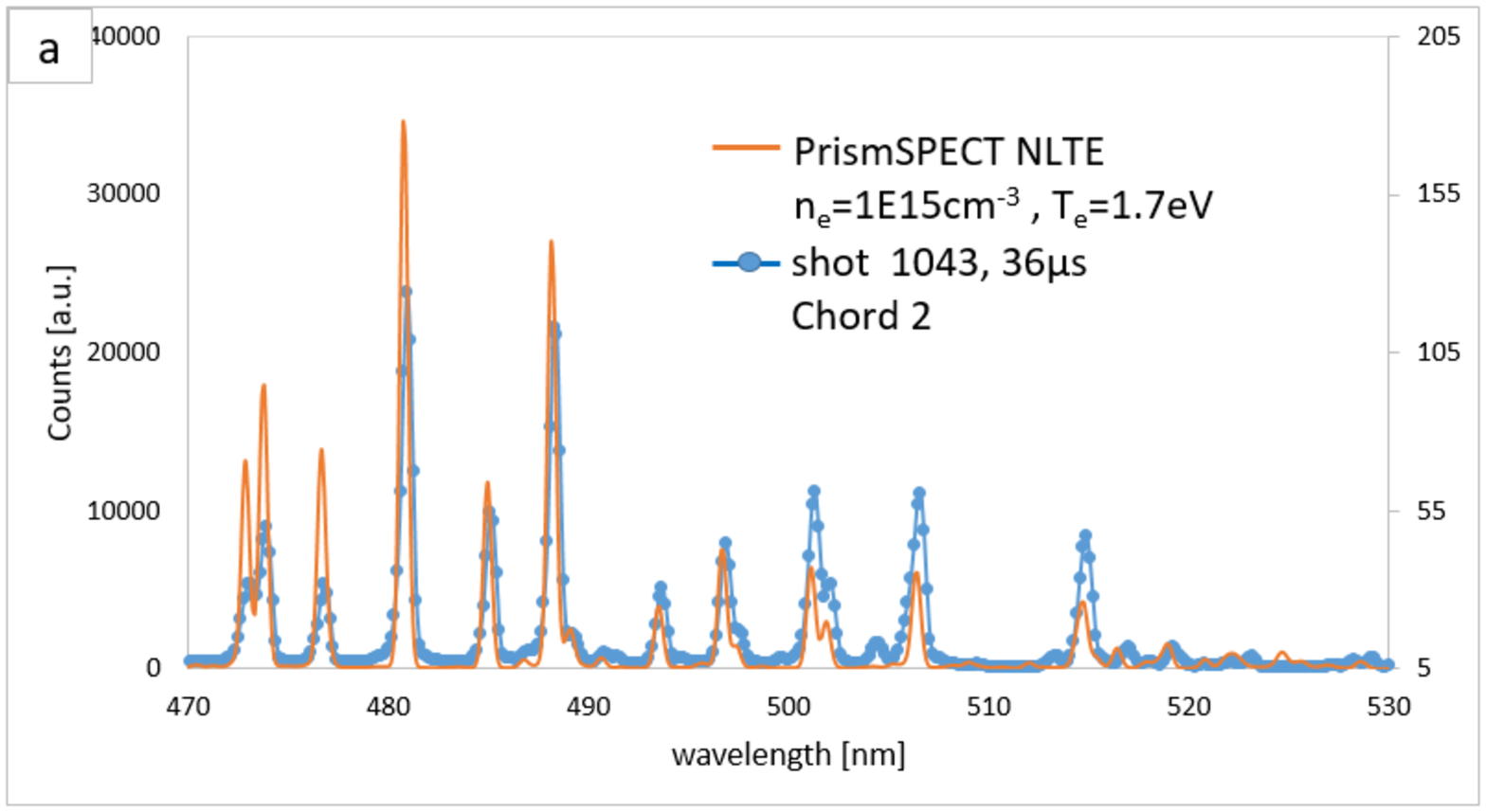}
\includegraphics[width=3.4truein]{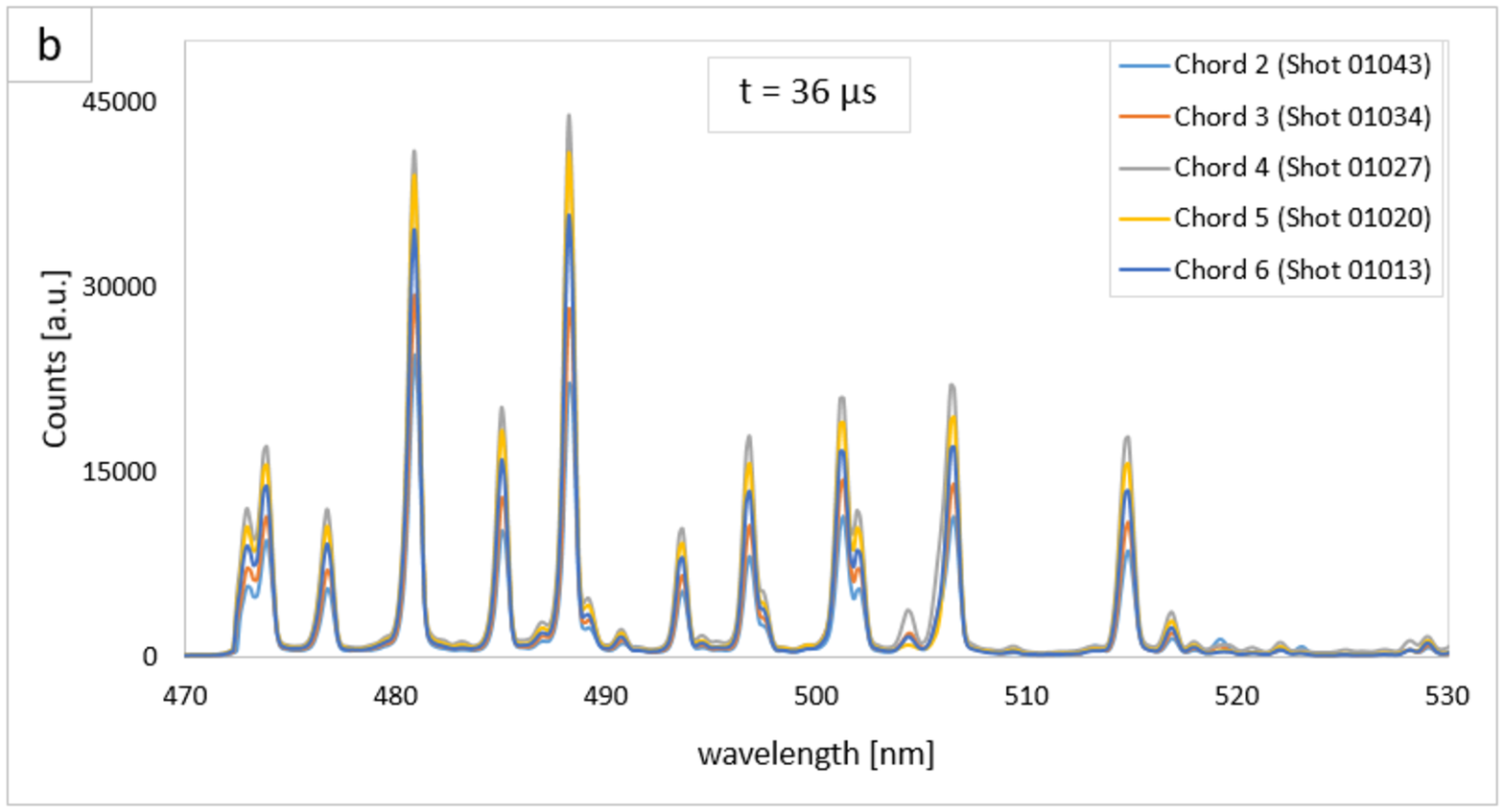}
\includegraphics[width=3.4truein]{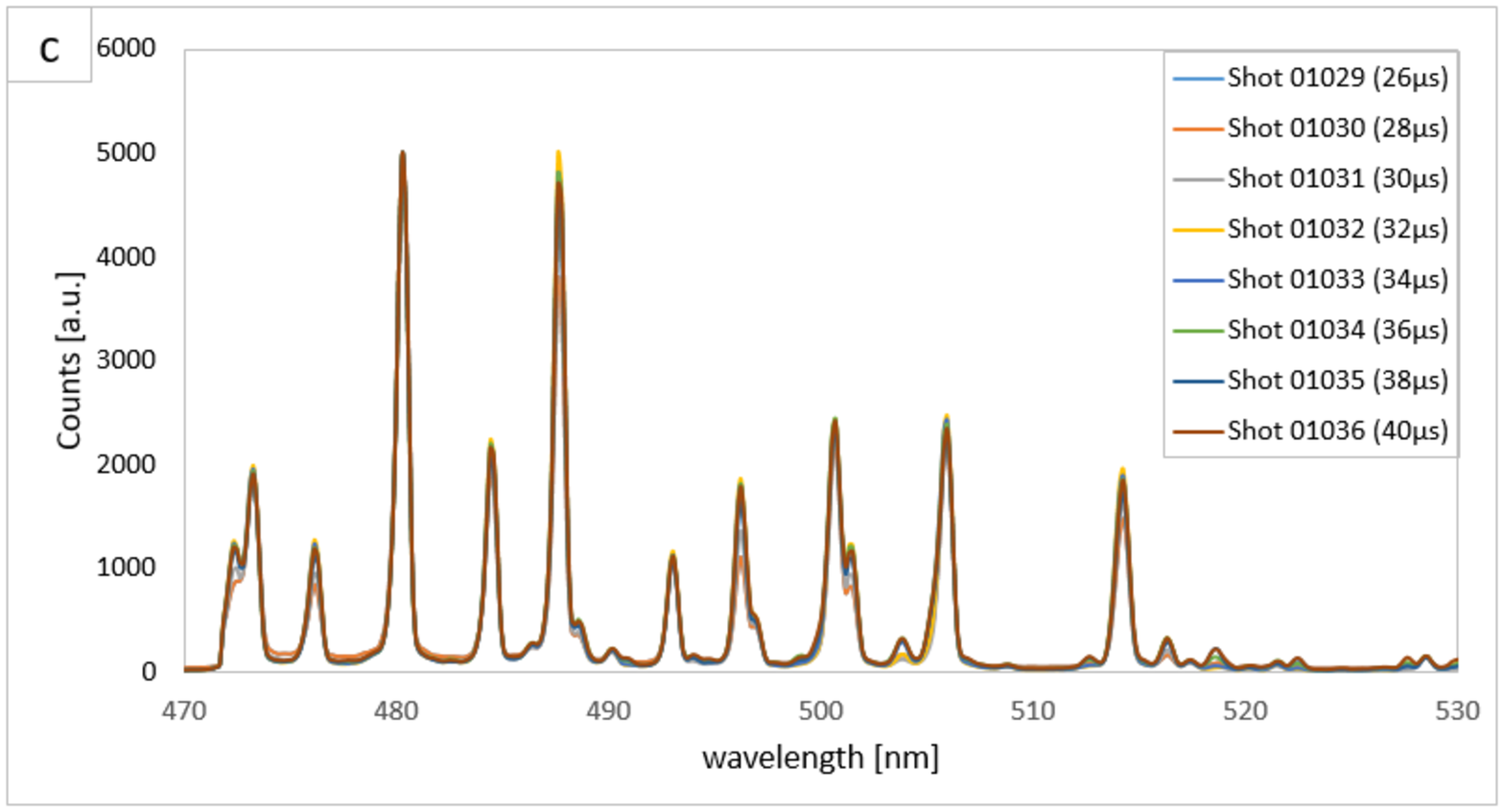}
\caption{\label{fig:Te} Survey-spectroscopy data for the chord positions and times shown in the plot legends, showing
that the spectra (corresponding to $T_e\approx 1.7$~eV and $\bar{Z}\approx 1.0$ based on
comparisons with PrismSPECT calculations) are largely unchanged over different 
positions and times.}
\end{figure}

Survey-spectroscopy data (Fig.~\ref{fig:Te}), along with $n_e$ from interferometry and comparisons with 
PrismSPECT\cite{macfarlane2004ifsa} calculations, are used to
estimate $T_e\approx 1.7$~eV and $\bar{Z}\approx 1.0$ (for $n_e=10^{15}$~cm$^{-3}$)
along the interferometer chord positions shown in Fig.~\ref{fig:6gundiag2}.
The results indicate very little variation over different chord positions and times, suggesting fairly uniform $T_e$
in the six-jet merging experiments.

\subsection{Morphology change by adding a seventh plasma jet}

Adding a seventh gun (to the middle of the hexagonal vertices where the six guns are mounted, see Fig.~\ref{fig:6gundiag2})
provides the opportunity to explore
the qualitative change in the structure of the liner section that is formed.  This also helps further
constrain and benchmark the simulations.  Figure~\ref{fig:iccd67gun} shows the different morphology
of the six- versus seven-jet merging cases.  As discussed briefly earlier in the paper, a new hexagonal pattern
of merging shocks appears due to the presence of the seventh jet.  Unfortunately, interferometry was not possible for
these experiments because the seventh gun uses the same chamber port as the interferometer launch optics.

\begin{figure}
\includegraphics[width=3.1in]{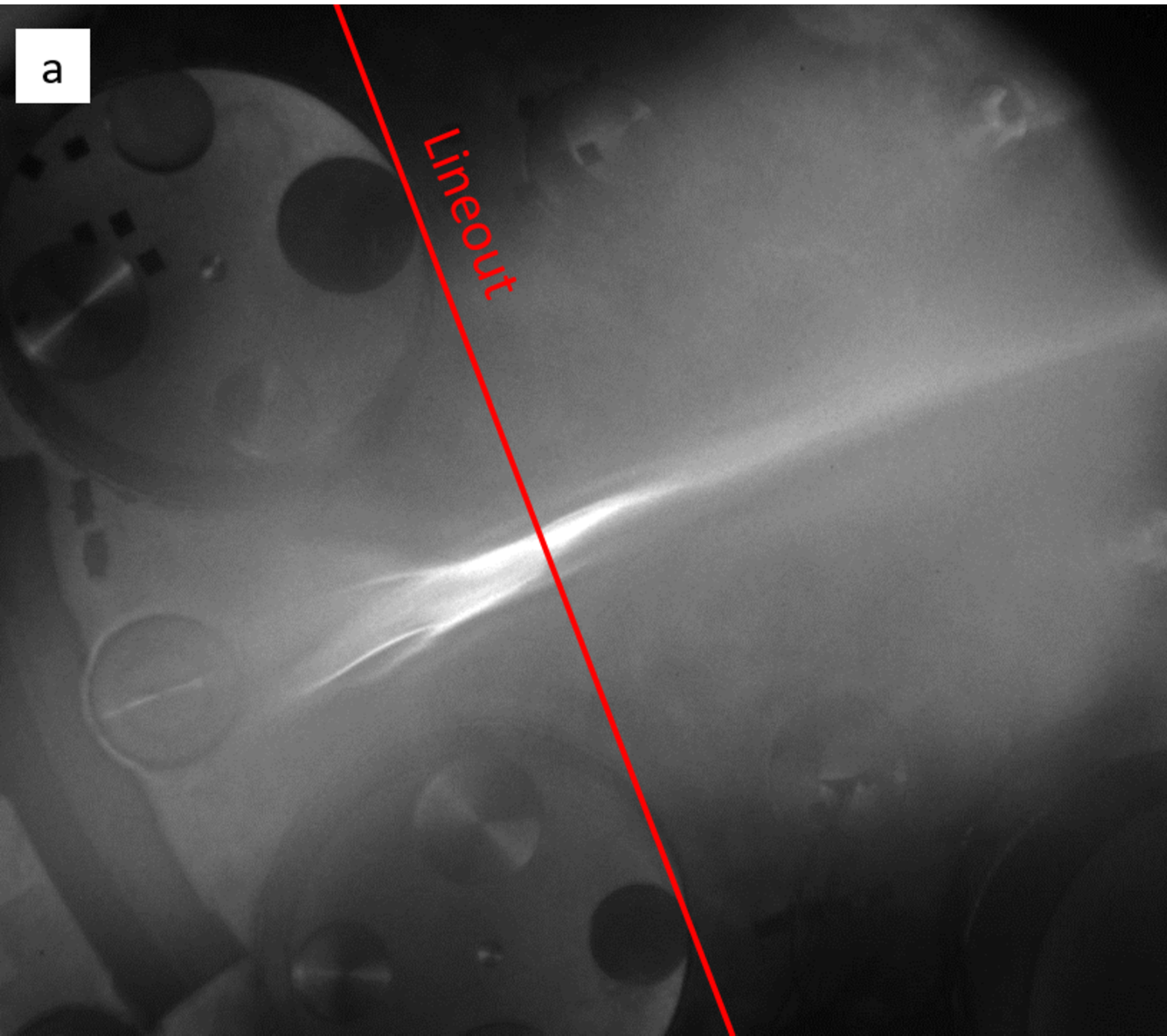}
\includegraphics[width=3.1in]{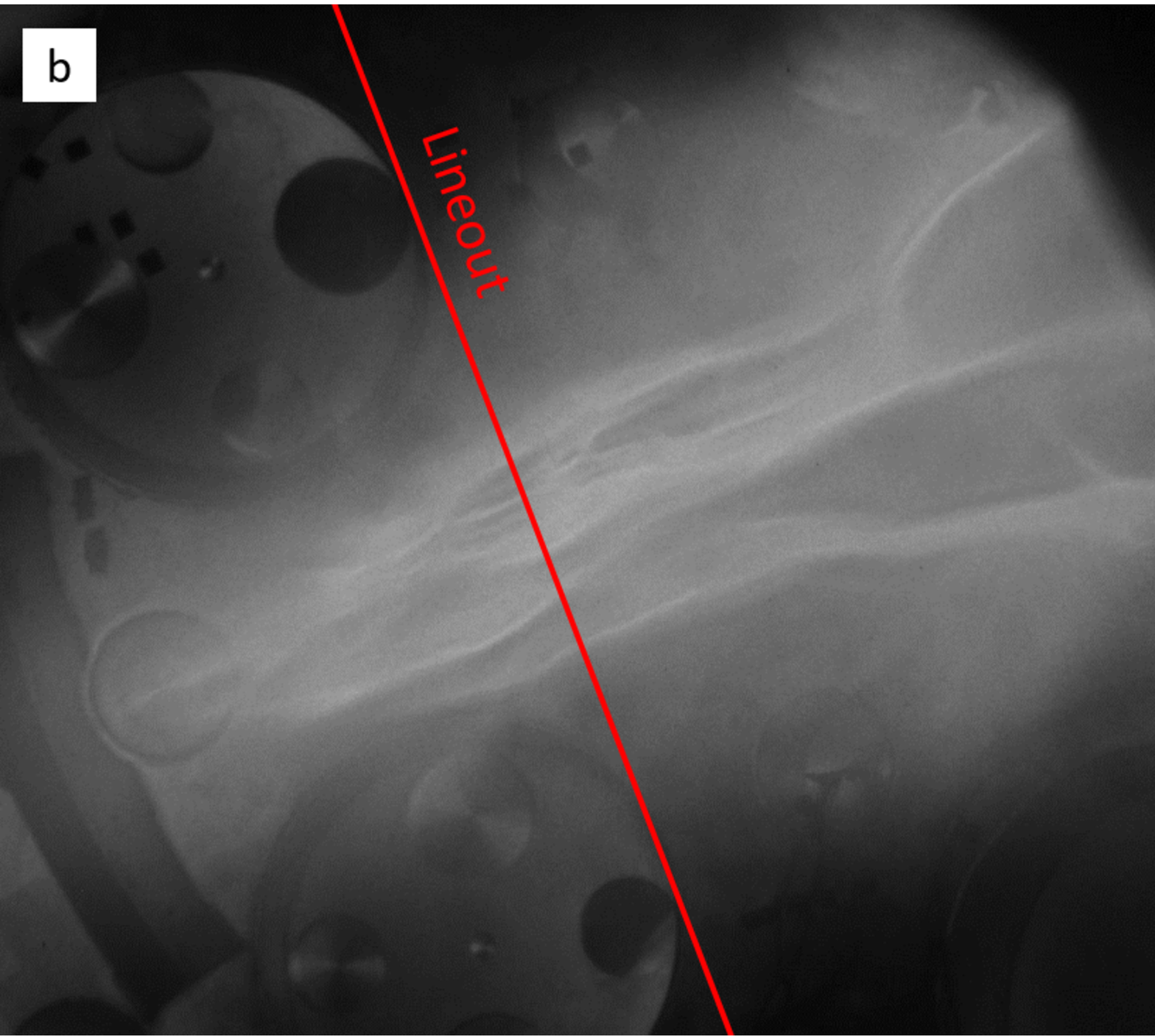}
\includegraphics[width=3.1in]{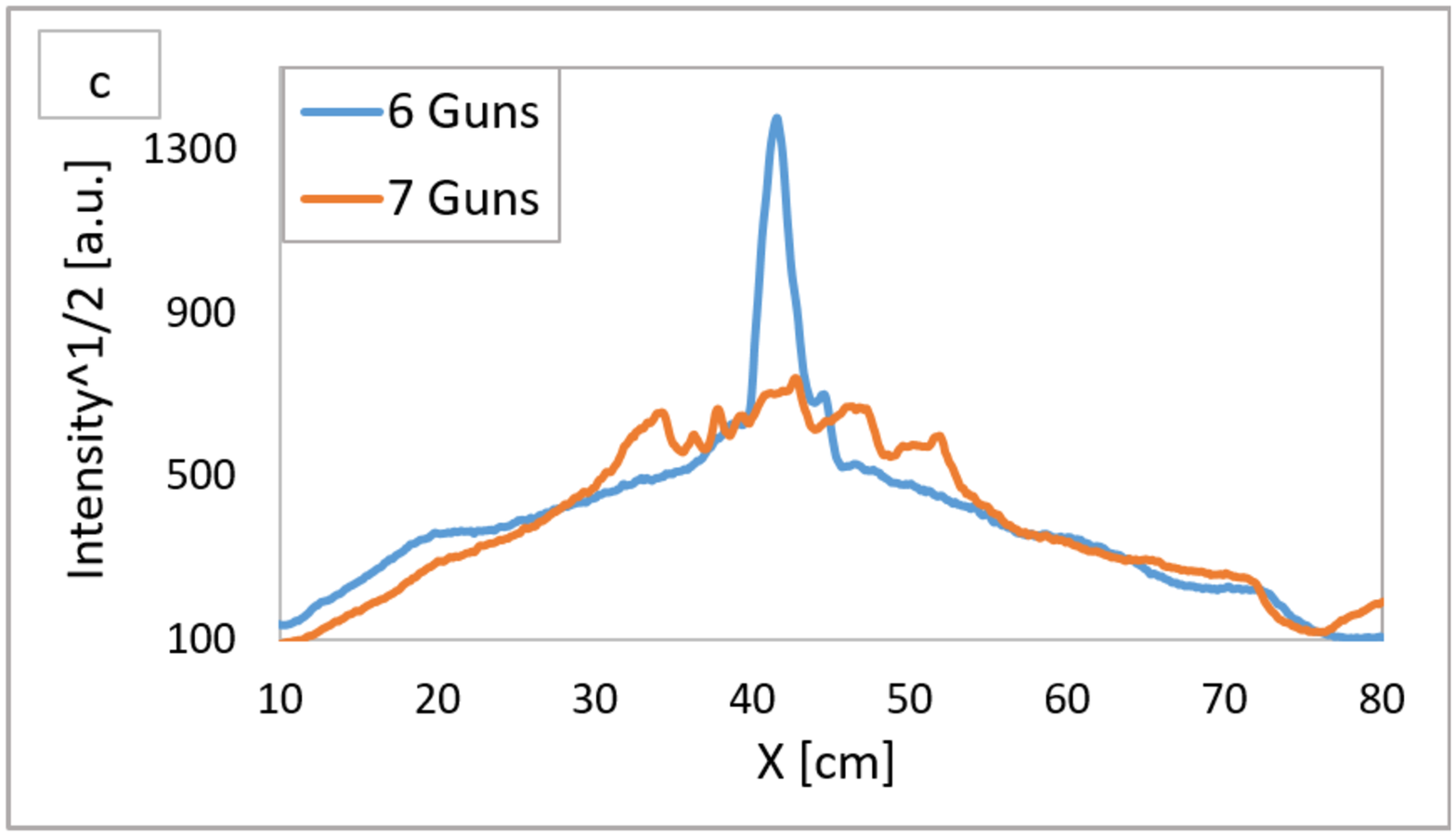}
\caption{Side-on, self-emission iCCD images (10-ns exposure, $t=34$~$\mu$s) of (a) six- and (b)~seven-jet merging.
(c)~Lineouts of the square root of the intensity from both images at approximately 20~cm from chamber center.}
\label{fig:6 and 7 gun comp}
\end{figure}

Side-on images show a dramatic change in shock structures with the introduction of the seventh jet, as seen in
Fig.~\ref{fig:6 and 7 gun comp}.  The six-jet case shows a sharp secondary shock formed by the merging of the primary
shocks along the symmetry axis of the six jets.  The seven-jet case
shows what appears to be several propagating primary shocks, presumably associated with the hexagonal
pattern seen in Fig.~\ref{fig:iccd67gun}(g).  The collisional merging of the 
six jets with the inner seventh jet modifies the propagation of the six jets as well as the six primary shocks, such that the 
secondary shock does not form at the same time and position as for the six-jet case.  Lineouts 
of the square root of image intensity across the merged structures are shown in Fig.~\ref{fig:6 and 7 gun comp}(c).
Assuming that $T_e$ is spatially uniform (as shown earlier in the paper), lineouts are representative of the
density profile.  The seven-jet case shows a much more uniform profile than the six-jet case, showing that
gun positioning and merging angles may be a way to optimize the liner uniformity.

\section{Summary and conclusions}
\label{sec:summary}

In this paper, we report comprehensive experimental results for the merging of multiple hypersonic jets to
form a section of a spherically imploding plasma liner.  This is a first major step toward developing plasma liners
as a standoff driver for MIF, based on the PJMIF concept.

The first experimental campaign (Sec.~\ref{sec:results1}) reported in this paper focused on 
the merging of two and three hypersonic plasma jets, as the most fundamental building block
of plasma-liner formation, to study and characterize plasma-shock
formation, spatial density non-uniformities, and liner Mach-number degradation. 
For plasma jets merging at larger half-angles (20.5$^\circ$), the merged plasma 
is more uniform due to the large interpenetration of the plasma jets and the softening or outright elimination of shock 
formation.  
For plasma jets merging at smaller half-angles (11.6$^\circ$), the merged plasma has larger density gradients due to the 
decrease in inter-jet interpenetration lengths, leading to plasma-shock formation.
Jet merging
is shown to slightly increase $T_e$ and $\bar{Z}$, leading to an increase in collisionality.  This effect, excacerbated by 
the higher $\bar{Z}$ of titanium impurities, increases collisionality in the merging process, resulting in shock formation and
density non-uniformities upon three-jet merging.  Upon jet merging, the Mach number degrades to about 4 due to shock 
heating of the ions but relaxes back to about 12 after $\sim$10~$\mu$s due to ion-electron equilibration.  Mach-number 
degradation does not appear to be a serious issue in this parameter space.  However, the formation of plasma 
shocks upon jet merging may pose an eventual threat to the uniformity of magnetized target compression.  
Obtaining this data is an important first step toward addressing the key question of what asymmetry is tolerable for PJMIF, where 
the number of jets, as well as their speed, merging angles, and axial profiles, all must be optimized to minimize liner
non-uniformity.

The second experimental campaign (Sec.~\ref{sec:results2}) reported in this paper
studied and characterized the formation of a section of a spherically imploding plasma liner formed
by six and seven plasma jets.  Upgraded GVs and a 
ballasting system were employed to achieve a significantly improved jet-to-jet mass balance of $<2$\% (see
Appendix~\ref{sec:jets}) compared to earlier work\cite{hsu2018experiment} and the results of Sec.~\ref{sec:results1}.
The improvements lead to a more balanced merging of six and seven plasma 
jets, which are also seen in benchmarked simulations from SPFMax and previously published results from 
FronTier.\cite{shih2019simulation} 
The uniformity of the merging process appears to be improved with the addition of a seventh gun in the middle of the
six guns.  Despite these improvements, interferometry measurements show appreciable
shot-to-shot variations in line-integrated electron densities, as well as imperfect mirror symmetry about the
symmetry axes.  These effects 
could, eventually, provide a seed for the Rayleigh-Taylor instability that lead to degraded compression of a magnetized plasma 
target.  Continued progress is
necessary to reduce the timing jitter (simulations\cite{shih2019simulation} suggest the need for $\leq$100~ns for
fusion relevance), the initial plasma-jet length, jet-to-jet velocity balance, and a reduction in shot-to-shot variations.  

Further gun upgrades
have been implemented since this work, with many of the new guns already mounted on the PLX chamber,
to improve on the aforementioned parameters and improve on gun robustness and maintainability.
The latest guns will be used in
upcoming 18- and 36-gun experiments to form hemispherical and fully spherical imploding plasma liners, respectively.
The goal of future 
experiments should include tuning the plasma-jet merging parameters, including reduction of impurities, to optimize the 
collisionality parameters such that the liner density is high enough to provide ion cooling with the surrounding electrons, 
yet have increased inter-jet penetration to soften or eliminate shock formation.
The experimental data presented in this paper are being utilized to benchmark multi-physics models and simulations to 
guide us in the optimization and design of future experiments.

\begin{acknowledgments}
This work was supported by the Advanced Research Projects Agency--Energy (ARPA-E) of the
U.S. Department of Energy (DOE) under contract no.\ DE-AC52-06NA25396 and cooperative agreement no.\ DE-AR0000566.
HyperJet Fusion Corporation acknowledges the support of Strong Atomics.
Original construction and operation
of the PLX facility at LANL and earlier development of contoured-gap coaxial plasma guns by
HyperV Technologies Corp.\ (both prior to 2012) were supported
by the DOE Office of Science, Fusion Energy Sciences.  The data that support the findings of this study are available from the corresponding author upon reasonable request.
\end{acknowledgments}

\appendix

\section{Additional data from merging of two and three jets (N, Kr, Xe)}
\label{sec:more_data}

\begin{figure}[tb]
\includegraphics[width=3.2truein]{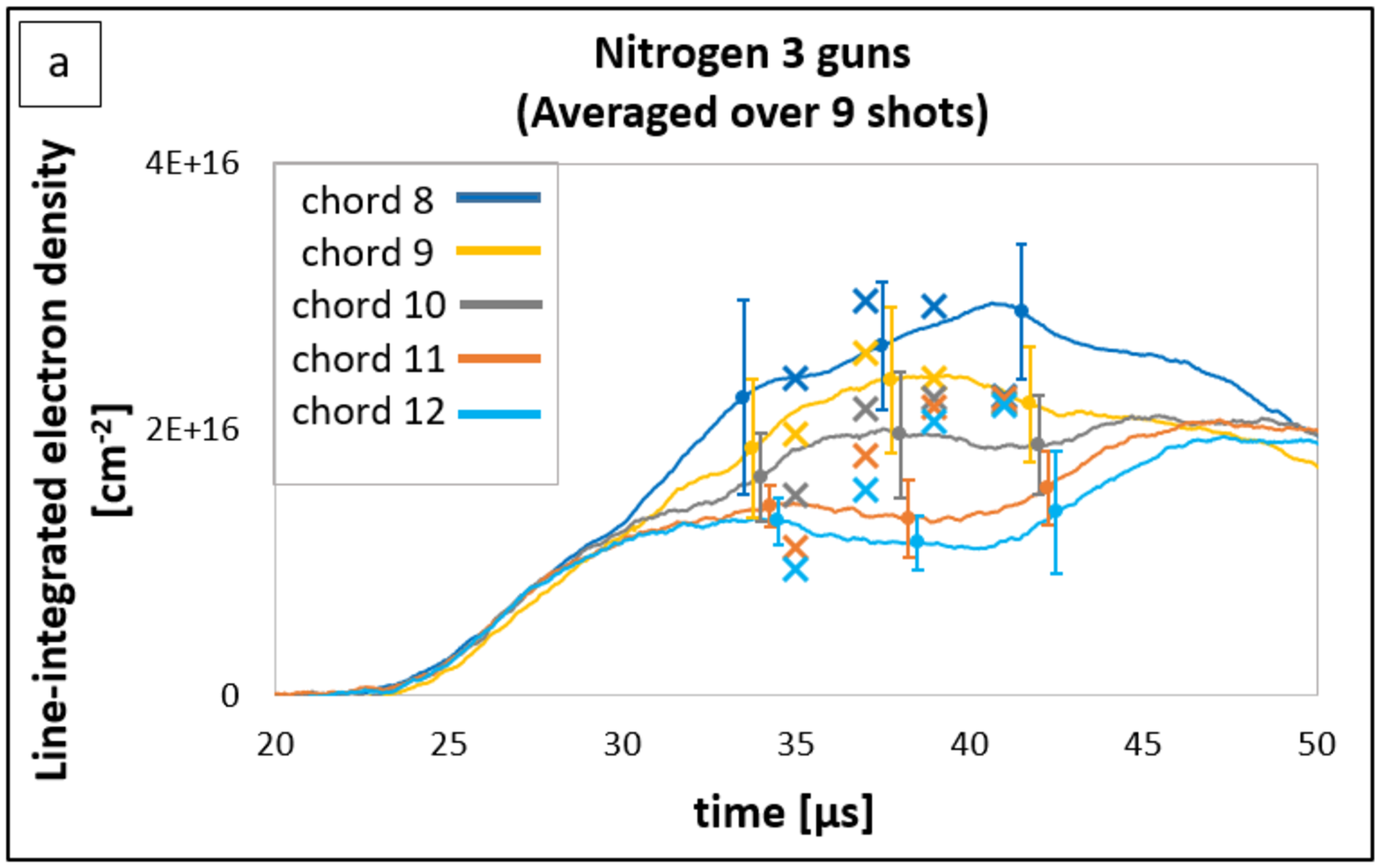}
\includegraphics[width=3.2truein]{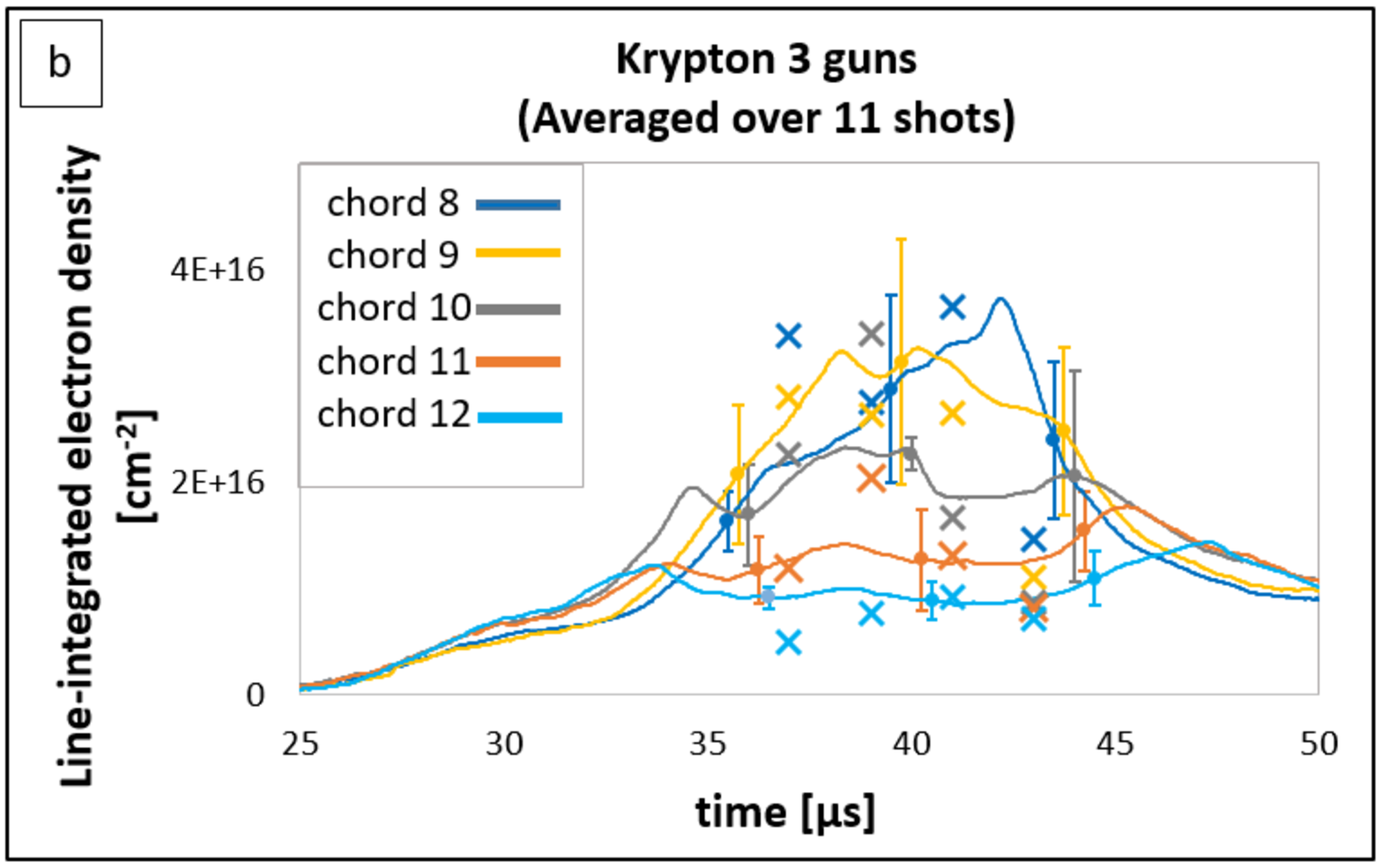}
\includegraphics[width=3.2truein]{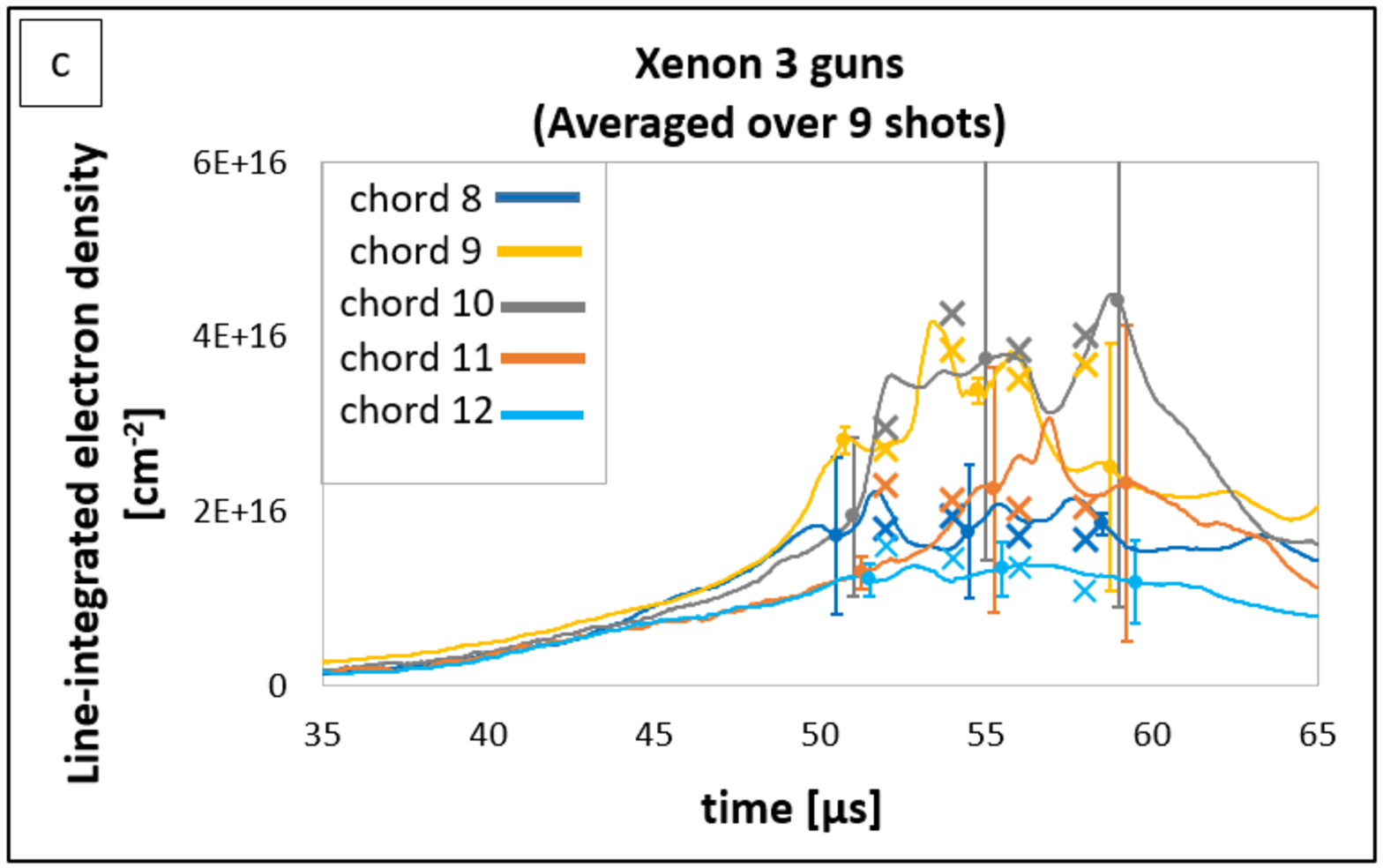}
\caption{\label{fig:NKrXeint} Interferometer signals versus time for the chord positions shown in Fig.~\ref{fig:diagnostics}
(green dots) for (a)~N, (b)~Kr, and (c)~Xe jets. See Sec.~\ref{sec:III.A} and the caption of Fig.~\ref{fig:gunsint2}.}
\end{figure}

\begin{figure}[tb]
\includegraphics[width=3.2truein]{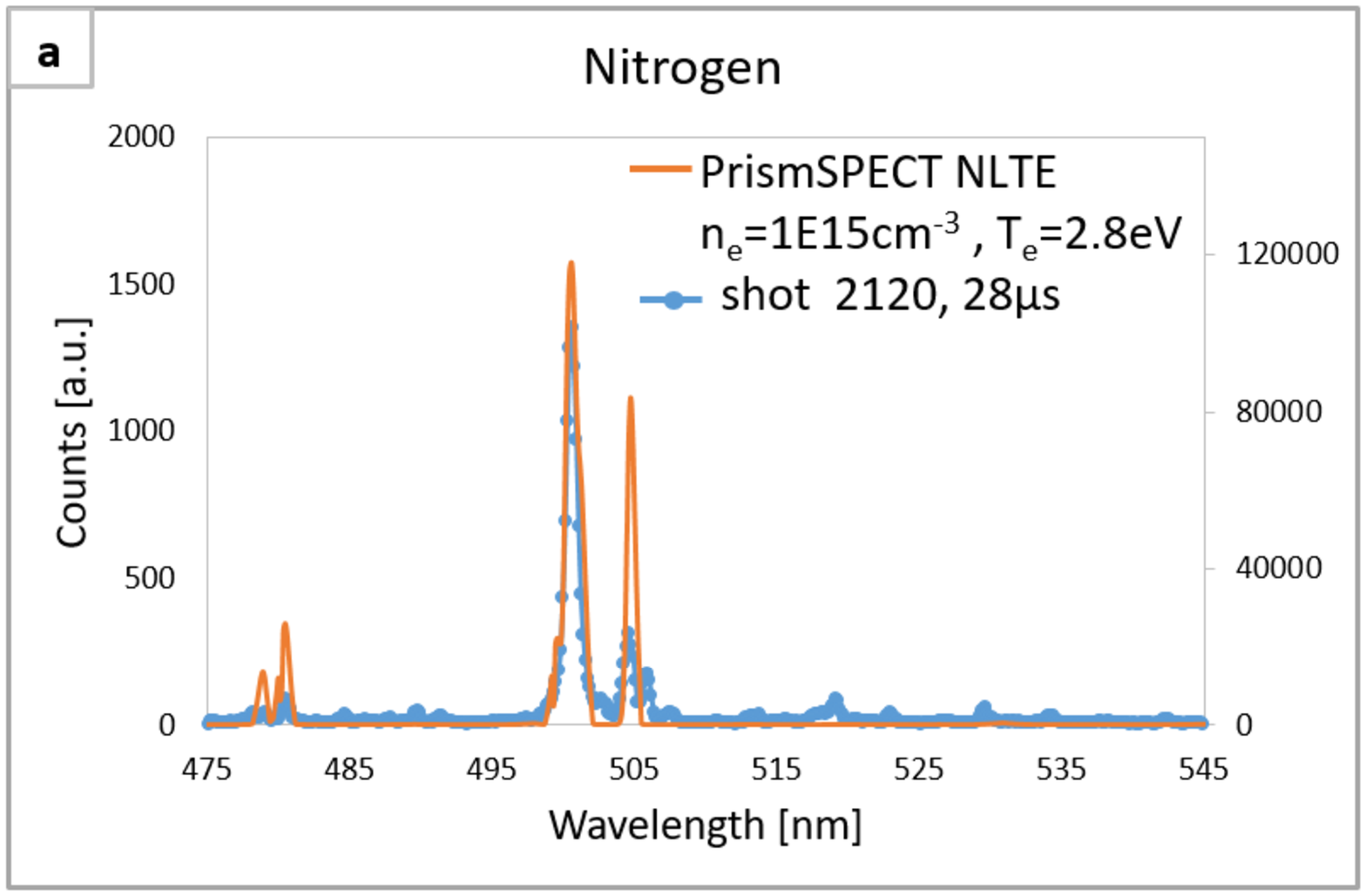}
\includegraphics[width=3.2truein]{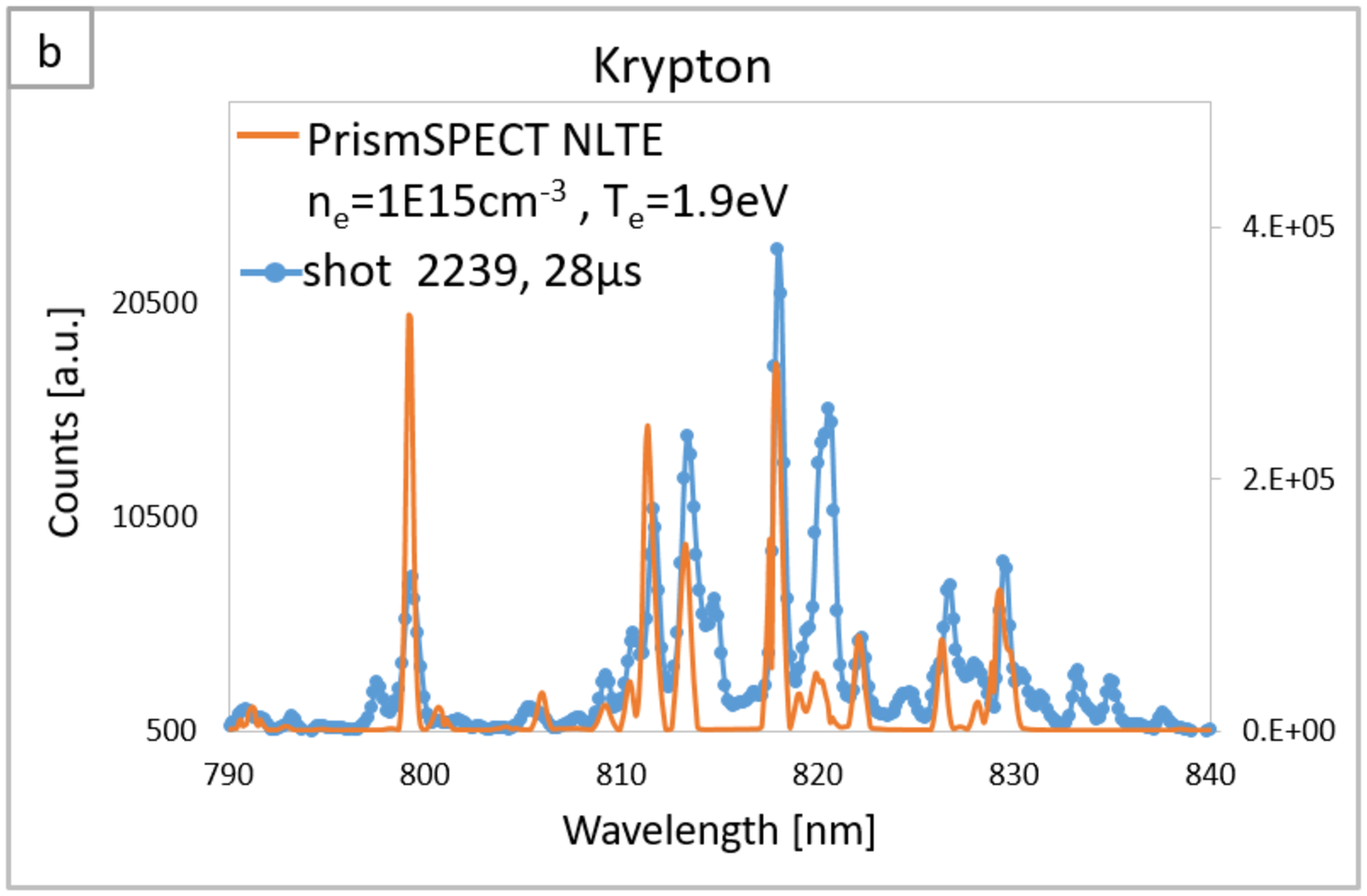}
\includegraphics[width=3.2truein]{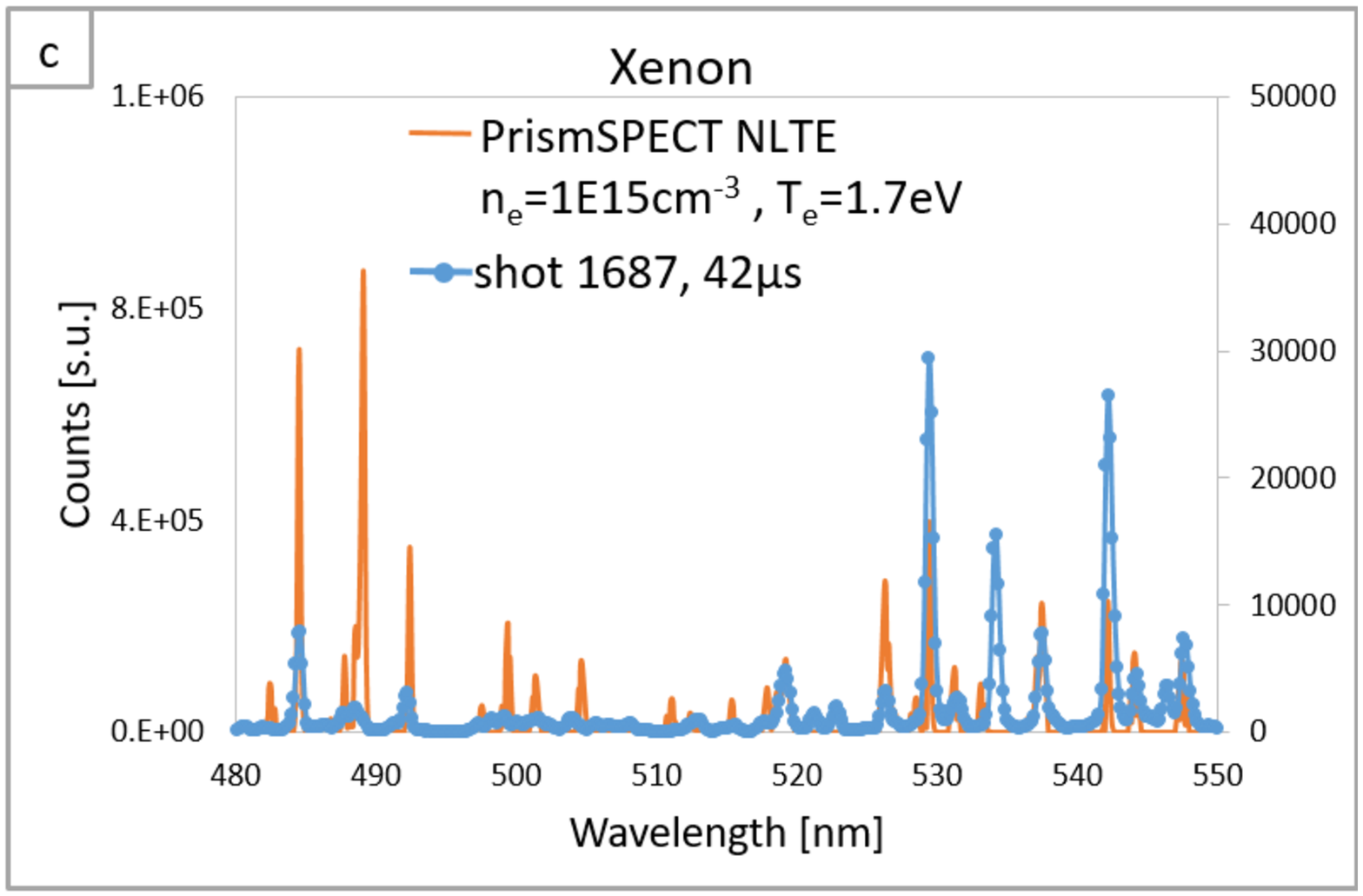}
\caption{\label{fig:NKrXespec} Comparison of experimental and calculated atomic spectra for three-jet
merging for N, Kr, and Xe, at the location of chord~10 (the middle
green dot in Fig.~\ref{fig:diagnostics}).  Inferred $n_e$ and $T_e$ are given in the legends of each panel.
See Sec.~\ref{sec:III.A}.}
\end{figure}

This appendix provides N, Kr, and Xe data, shown in Figs.~\ref{fig:NKrXeint} and \ref{fig:NKrXespec}, supplementing
the Ar data shown in Figs.~\ref{fig:gunsint2} and \ref{fig:nitargspec2}, respectively.  See the discussion in Sec.~\ref{sec:III.A}.

\section{Improved jet-to-jet balance}
\label{sec:jets}

The results presented in Sec.~\ref{sec:results2} are based on improved jet-to-jet balance achieved
via upgraded GVs and fine-tuned series (ballast) GV impedances.  This appendix presents detailed
experimental characterization of the improved jet-to-jet balance.

The first six-jet merging experiments on PLX revealed that jet-to-jet mass imbalance ($\gtrsim 20$\% variation across jets)
degraded the
expected mirror symmetry in density profiles and merging morphology about the jet-propagation
axes.\cite{hsu2018experiment} Thus, we made significant improvements
to the GVs in the plasma guns to reduce the mass variation among jets.  
The earlier generation of GVs consisted of an in-line plenum, magnetic coils, a flyer plate, and eight metal springs.
The flyer plate rests on the eight uncompressed springs.  When the GV coils are pulsed with a 10-kV 
capacitor bank, magnetic pressure pushes the flyer plate against the springs, compressing them far enough to allow the flyer 
plate to be pushed below gas-inlet holes at the rear of the gun and letting in pressurized gas. 
In the new GVs, a modified plenum increases precision in the plenum volume.  It also increases the volume, now with an
annular profile and a smaller cross-sectional area.  This allows higher plenum pressures and larger dispensed masses.  
Previous GVs applied pressure to the entire flyer plate, whereas the new GV applies annular pressure to the outer rim of the flyer 
plate.  The flyer plate now has an additional fin to direct gas flow towards the inlet holes to the gun breech rather than towards
the center of the plate.  

\begin{figure}[!tb]
\includegraphics[width=2.8truein]{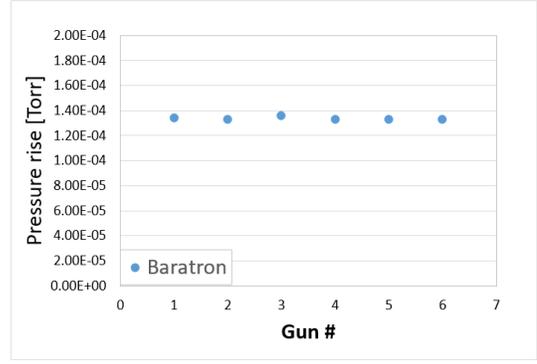}
\caption{\label{fig:jetmass2} Chamber pressure rise measured by a baratron pressure gauge
for gas injection from each of six plasma guns with upgraded GVs, 
after fine tuning of ballast inductances and resistances in 
series with each GV\@.  Results show 1.7\% standard deviation in mass injected across six guns.}
\end{figure}

However, even with the improved GV design, further fine tuning was still
required especially because all the GVs are driven in parallel
by a single capacitor bank.  The fine tuning is enabled via a ballasting system that
allows for adjusting the inductances and resistances in series with each GV to control the split of electrical current delivered to
each GV.  The fine-tuning sequence is as follows.
Gas lines are disconnected from all but one GV and evacuated.  The mass injection by the one GV is determined by monitoring the increase in vacuum chamber pressure when all GVs are pulsed.  The increase in chamber pressure is 
determined using an MKS Instruments model 626C.1TLF baratron pressure gauge.  
The ballast inductance and resistance is adjusted to arrive at an injected mass of 
4~mg.  This process is repeated for each different GV.  Another iteration or two through all GVs is sometimes needed.
Using the modified GVs followed by the tuning of ballast inductances and resistances
results in $<2$\% variation in the injected mass across all guns, as shown in Fig.~\ref{fig:jetmass2}.

\begin{figure}[!tb]
\includegraphics[width=2.8truein]{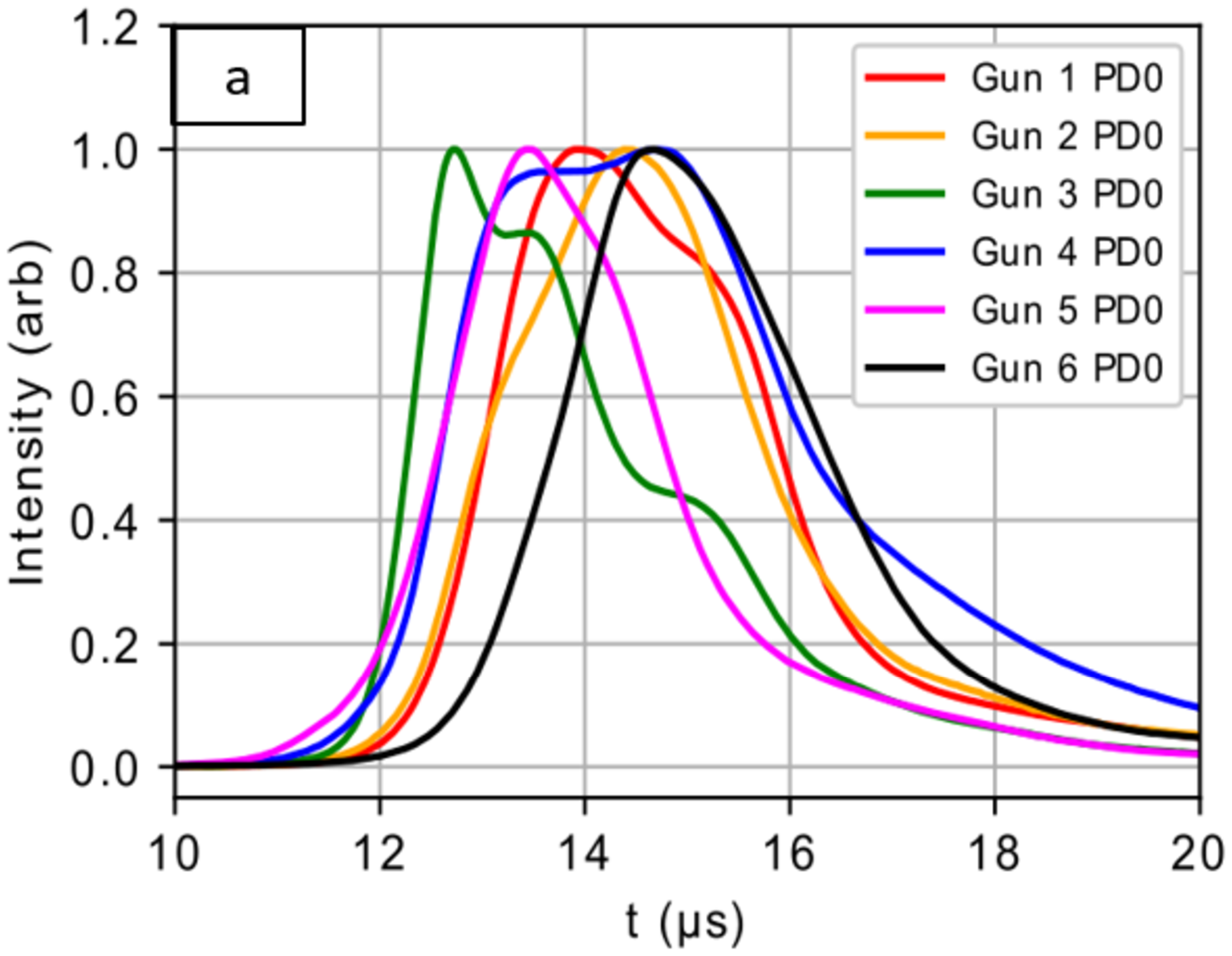}
\includegraphics[width=2.8truein]{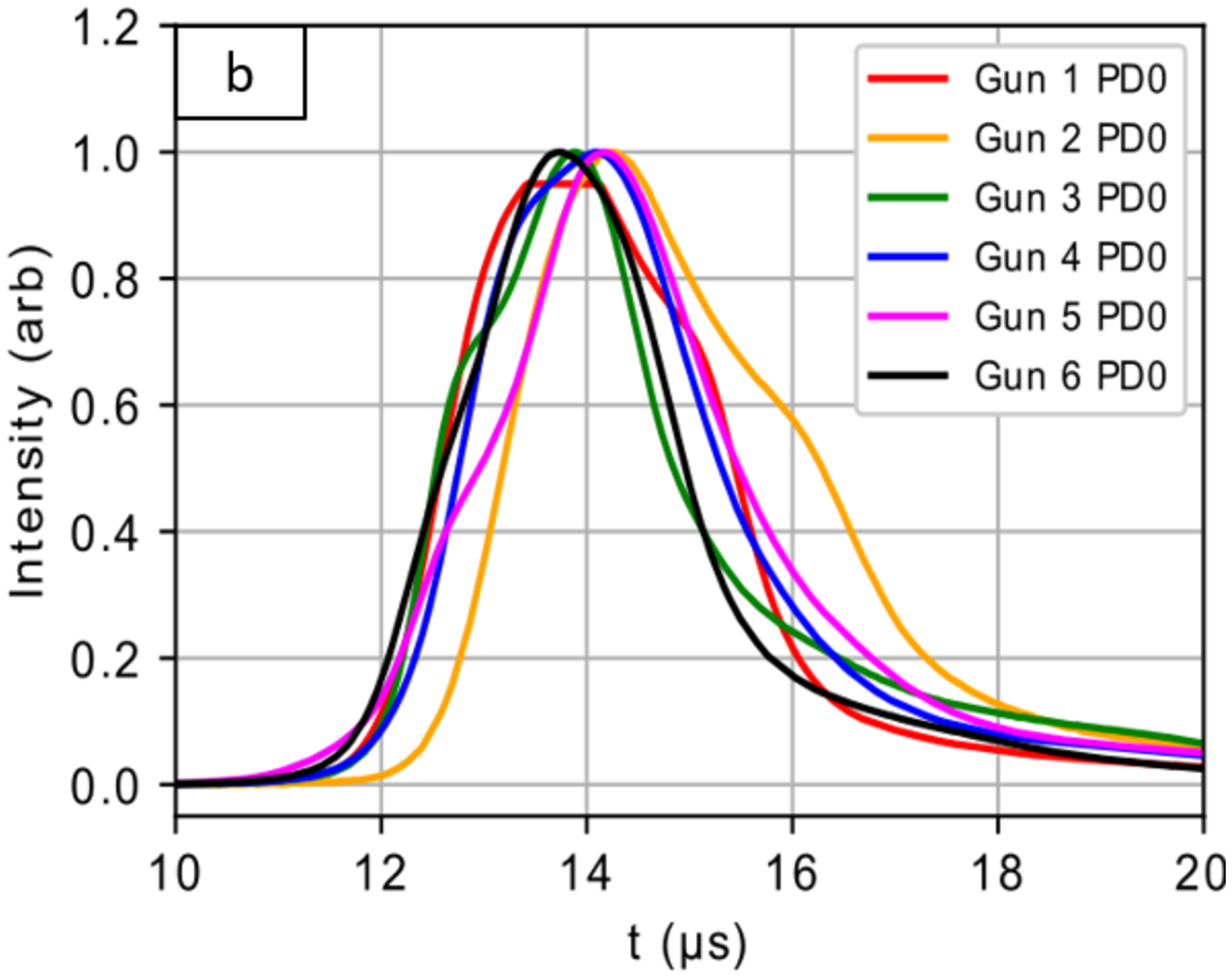}
\caption{\label{fig:pdbadgood} Photodiode (PD) traces of six jets at the exit of their gun nozzles (see Fig.~\ref{fig:gun} 
for photodiode viewing chords) for the upgraded GVs,
(a) before and (b) after fine tuning of ballast inductances and resistances.}
\end{figure}

The improved synchronicity of initial jet propagation across all jets is verified via photodiode data (Fig.~\ref{fig:pdbadgood})
of the jet propagation
as it exits the gun muzzle.  Each gun has two photodiode views, separated by 2~cm, transverse to the jet-propagation axis
near the exit of the gun nozzle (see Fig.~\ref{fig:gun}).
The timing jitter for the six guns is determined by measuring the spread in the photodiode signals.
The arrival time of a jet is taken to be the time that photodiode signal reaches half its maximum.
Data is analyzed for ten shots, showing an average variation in the arrival time of 610~ns with a standard deviation of 200~ns.  
Figure~\ref{fig:pdbadgood} shows photodiode traces before and after the GV/ballast improvements.  
Shot-to-shot jitter in the injected mass now sets the lower bound in the timing variation across jets.  End-on
imaging of six- and seven-jet merging (e.g., Fig.~\ref{fig:iccd67gun}) quickly alerts us
if one of the guns is firing with mismatched mass and/or timing.

\begin{figure}[!tb]
\includegraphics[width=3.4truein]{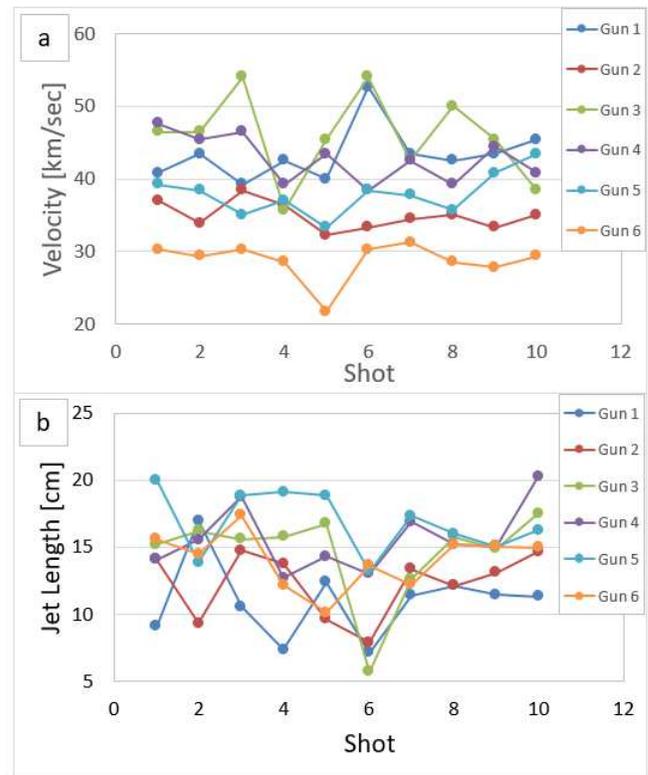}
\caption{\label{fig:lengthvel} Jet velocities and lengths from all six plasma guns with upgraded GVs, inferred from 
photodiode data.}
\end{figure}

The photodiode data are also use to determine the velocity and length of the jets (Fig.~\ref{fig:lengthvel}).
The velocity is determined from the difference in arrival times of the two photodiode signals for each gun.  The length
is from the full-width, half-maximum of the photodiode signals.
The jet velocities are between 25 and 55~km/s, with the jet from gun three being relatively fast and from gun six being relatively
slow.
The lengths of the plasma jet upon exiting the nozzle are in the range 7--20~cm.  The variations in timing, velocity, and jet length 
across the guns, as well as from shot-to-shot jitter, show that continued engineering improvements in the gun are needed.

%

\end{document}